\documentclass[a4paper,fleqn,usenatbib]{mnras}

\usepackage{newtxtext,newtxmath,times}

\usepackage[T1]{fontenc}
\usepackage{ae,aecompl}

\usepackage{subfig}
\usepackage{graphicx}	
\usepackage{amsmath}	
\usepackage{amssymb}	
\usepackage{array}
\usepackage{physics}
\usepackage{tabularx}

\newcommand{\rsun}{R$_\odot$}
\newcommand{\msun}{{M$_\odot$}}
\newcolumntype{x}[1]{>{\centering\arraybackslash}p{#1}}
\newcolumntype{Y}{>{\centering\arraybackslash}X}
\let\vec\mathbf

\title[Recombination energy in common envelope interactions]{The impact of recombination energy on simulations of the common envelope binary interaction}
\author[T. A. Reichardt et al.]{Thomas A. Reichardt$^{1,2}$\thanks{E-mail: thomas.reichardt@students.mq.edu.au},
Orsola De Marco$^{1,2}$, 
Roberto Iaconi$^{4,1,2}$\thanks{JSPS International Research Fellow (Graduate School of Science, Kyoto University)},
\newauthor and Daniel J. Price$^{3}$ \\
$^{1}$Department of Physics and Astronomy, Macquarie University, Sydney, NSW 2109, Australia \\
$^{2}$Astronomy, Astrophysics and Astrophotonics Research Centre, Macquarie University, Sydney, NSW 2109, Australia \\
$^{3}$School of Physics and Astronomy, Monash University, VIC 3800, Australia\\
$^{4}$Department of Astronomy, Kyoto University, Kitashirakawa-Oiwake-cho, Sakyo-ku, Kyoto 606-8502, Japan 0000-0002-1940-1950\\
}

\date{Accepted XXX. Received YYY; in original form ZZZ}

\pubyear{2018}

\begin{document}
\label{firstpage}
\pagerange{\pageref{firstpage}--\pageref{lastpage}}
\maketitle

\begin{abstract}
  During the common envelope binary interaction, the expanding layers of the gaseous common envelope recombine and the resulting recombination energy has been suggested as a contributing factor to the ejection of the envelope. In this paper we perform a comparative study between simulations with and without the inclusion of recombination energy. We use two distinct setups, comprising a 0.88-\msun\ and a 1.8-\msun\ giants, that have been studied before and can serve as benchmarks. In so doing we conclude that (i) the final orbital separation is not affected by the choice of equation of state. In other words, simulations that unbind but a small fraction of the envelope result in similar final separations to those that, thanks to recombination energy, unbind a far larger fractions. (ii) The adoption of a tabulated equation of state results in a much greater fraction of unbound envelope and we demonstrate the cause of this to be the release of recombination energy. (iii) The fraction of hydrogen recombination energy that is allowed to do work should be about half of that which our adiabatic simulations use. (iv) However, for the heavier star simulation we conclude that it is helium and not hydrogen recombination energy that unbinds the gas and we determine that all helium recombination energy is thermalised in the envelope and does work. (v)~The outer regions of the expanding common envelope are likely to see the formation of dust. This dust would promote additional unbinding and shaping of the ejected envelope into axisymmetric morphologies.
\end{abstract}




\section{Introduction}

A common envelope interaction \citep{paczynski1976common, ivanova2013common} occurs when the orbital separation between two stars decreases to the extent that they come to share the same envelope. The existence of compact, evolved binaries implies that in at least a fraction of all common envelope interactions, the envelope is fully ejected. Yet, hydrodynamic simulations that utilise an ideal equation of state do not unbind a sufficient fraction of the envelope to allow us to conclude that the binary will survive \citep[e.g.,][]{reichardt2019extending}.

\citet{nandez2015recombination} and \citet{ivanova2016common} simulated common envelope interactions between a 1.8\,\msun\ star, and a range of companions with masses between 0.05 and 0.36\,\msun\ by using a tabulated equation of state that captures the effects of recombination energy. \citet{ivanova2016common} reported that, with a companion mass of 0.36\,\msun, approximately half of the envelope is ejected by 50\,d after the end of the fast inspiral. The remaining bound envelope material is subsequently ejected over a timescale of approximately 700\,days after the end of the fast inspiral. This later ejection is presumably at the hand of the recombination energy released, which is immediately and completely thermalised thanks to the adiabatic approximation adopted by those simulations. \citet{nandez2015recombination} also noted that the same simulation unbound only about half of the envelope when utilising an ideal gas equation of state with radiation pressure, but a comparison between simulations with and without recombination energy was not presented by them.

The idea that the photons released during recombination could be used as an energy source to help with envelope unbinding was originally suggested by \citet{lucy1967formation} and \citet{roxburgh1967origin}. \citet{han1994possible} and \citet{harpaz1998role}, in particular, considered the expanding envelopes of pulsating AGB stars and whether recombination energy could facilitate the ejection of the envelope in these single stars, leading to the formation of planetary nebulae. \citet{harpaz1998role}, however, argued that recombination energy cannot be used in the ejection of the AGB envelope, because the envelope becomes transparent when it recombines, hence the energy will be transported out immediately by radiation.

Along similar lines of reasoning, \citet{sabach2017energy}, \citet{grichener2018limited} and \citet{soker2018radiating} utilised 1D common envelope simulations, carried out in \textsc{mesa}, to argue that only a fraction of the energy released by recombination of hydrogen and helium can be used to eject the common envelope, as the energy will be transported out through photon diffusion and convection. They estimated that only approximately 20~percent of the recombination energy can be utilised in the ejection of the common envelope, by comparing the timescales of envelope expansion and energy transport.

\citet{ivanova2018use} countered that the majority of recombination energy can be used to eject the envelope. They reasoned that the ratio of the radiation flux to the convective flux is small in most parts of the envelope, hence there is little reason to consider the effect of photon diffusion. Rather, it can be shown that the dominant transport mechanism in the region where recombination energy is released is actually convection. They further argued that even the maximum convective flux becomes very inefficient for transporting recombination energy in regions where the ionisation fraction is about 0.2. As a result, they concluded that the majority of the recombination photons will not be transported out of the envelope, hence they are destined to be thermalised and able to do work, increasing the expansion rate of the envelope. 

In order to resolve the debate of how much recombination energy can be used to do work as opposed to how much should be radiated away, a full treatment of radiation transport in common envelope simulations should be implemented. Some work in this direction is being carried out \citep{zhu2019recombination}. This line of work presents considerable challenges, chiefly because of the wall clock time required for these simulations (weeks to months \textit{without} including radiation transport). 

In this work we perform several simulations to quantify some of the effects of recombination energy, with the aim of determining how it should be used in 3D adiabatic codes in the absence of a reasonable treatment of radiation transport (as is the case for \citealt{ivanova2016common} and \citealt{nandez2016common}). We start by investigating {\it where and when} the gas is unbound with respect to where and when the recombination energy of both hydrogen and helium is released. 

We present two sets of simulations. The first set uses the initial conditions utilised by \citet{passy2012simulating}, but also studied in \citet{iaconi2017effect, iaconi2018effect} and \citet{reichardt2019extending}. This provides a well studied setup to which we can compare the effects of recombination energy. The second set is a direct comparison with the work of \citet{nandez2016common}, so as to calibrate our work and technique to theirs and ensure that our conclusions are compatible. Each set contains two simulations carried out with an ideal gas equation of state and two with a tabulated one based upon the equation of state used in the 1D stellar evolution code \textsc{mesa} \citep{paxton2011mesa}. In this way we have a direct comparison of the unbinding dynamics.

In Section~\ref{sec:simulations}, we describe the simulation setup, giving details of the tabulated equation of state. In Section~\ref{sec:results}, we describe the results of the simulations, including giving results of resolution and other numerical tests. After this, in Section~\ref{ssec:where_recombination}, we analyse the location of the delivery of recombination energy and compare it with where particles are being unbound (Section~\ref{ssec:where_unbound}), and where recombination energy is being released (Section~\ref{ssec:recombination_release}). In Section~\ref{ssec:can_it_be_used}, we move on to testing whether or not the use of the recombination is physically realistic. In Section~\ref{ssec:dusty_shell}, we discuss the apparent emergence of a dusty shell in the ejecta of the common envelope interaction. We summarise and conclude in Section~\ref{sec:mesa_conclusions}.

\section{Simulations}
\label{sec:simulations}
We performed a series of four simulations of the common envelope interaction. We used two different primary stars in these simulations. The first was a 0.88\,\msun\ RGB star with a 0.39\,\msun\ core and initial radius of about 80\,\rsun, used also in the simulations of \citet{passy2012simulating}, \citet{iaconi2017effect} and \citet{reichardt2019extending}. This star will henceforth be referred to as the P12 star. The second stellar setup had a greater initial mass of 1.80\,\msun, with a 0.32\,\msun\ core and an initial radius of 16\,\rsun. This second star, which we will refer to as the N16 star, is one of the setups used in the common envelope simulations of \citet{nandez2015recombination}, \citet{ivanova2016common} and \citet{nandez2016common}. We chose this star because \citet{ivanova2016common} note that it unbinds the entirety of the envelope. We have, in fact, utilised the exact stellar structure kindly provided by that group, so as to minimise the difference between the two simulations.

The P12 input stellar profile was evolved in the one-dimensional stellar evolution code \textsc{mesa} \citep{paxton2011modules}. The N16 star was also evolved with \textsc{mesa} by \citet{nandez2015recombination}. Before mapping these profiles into the computational domain, we use the procedure laid out by \citet{ohlmann2016hydrodynamic} to produce more stable giant stars in hydrodynamic simulations. For each star, we choose a radius from the centre (6\,\rsun\ for P12, 2\,\rsun\ for N16), at which we cut the stellar profile. Within this radius, the profile is recreated with a modified Lane-Emden equation using an index $n=3$, which is set to transition smoothly to the original profile at the cutoff radius. This modified Lane-Emden equation includes a contribution from a softened gravitational potential typically used in the core of a simulated giant star.

The new 1D profiles were then mapped into the 3D computational domain of the smoothed particle hydrodynamics code \textsc{phantom} \citep{price2017phantom}, with all stars containing approximately $1.4 \times 10^5$ SPH particles. A point mass particle is placed in the centre of the star, with a gravitational softening length equal to half of the cutoff radius for the modified profile. Therefore, for the P12 star, the core had a mass of 0.39\,\msun\ and a softening length of $h_\text{soft} = 3$\,\rsun. However, for the N16 star, the softening length was $h_\text{soft} = 1$\,\rsun, larger than the 0.15\,\rsun\ softening length used by \citet{ivanova2016common}. As the masses of the core particles are set to be consistent with the modified Lane-Emden profile, they may differ slightly from the desired value. Our core particle mass of 0.320\,\msun is slightly larger than the 0.318\,\msun\ of \citet{ivanova2016common}.

Our choice of softening length was dictated by prohibitively long computational times associated with small softening lengths. However, we did perform simulations with a smaller softening length of $h_\text{soft}=0.5$\,\rsun\ for both companion and core particles, with an associated primary core mass of 0.313\,\msun\ (see Section~\ref{ssec:movement_out_of_plane}).

The velocities in our stars were damped over five dynamical timescales, after which the stars were evolved with no damping in the computational domain for another five dynamical timescales. During this time they showed no significant expansion, proving that our stars are sufficiently stable.

Our SPH particles were assumed to have a constant chemical composition, defined by the hydrogen, helium and metal mass fractions. The compositions were taken from the atmospheres of the input stellar profiles, calculated by \textsc{mesa}. For the P12 star, the composition is set to be $X=0.67$, $Y=0.31$ and $Z=0.02$, and for the N16 star, we have $X=0.68$, $Y=0.30$ and $Z=0.02$. These compositions were used as inputs to the tabulated equation of state used by \textsc{mesa}, which we utilised in two of our four simulations (see Section~\ref{ssec:eos} below for details).

A companion star was then initialised in the computational domain as a second point mass particle, with a softening length equal to that of the core of the primary star. For the P12 simulations, the initial orbital separation is set to 100\,\rsun, which is just greater than the initial radius of the primary star. The N16 simulations were initialised with an orbital separation of 31.37\,\rsun, matching the initial separation used by \citet{nandez2016common}. We will adopt the notation referring to the ideal and tabulated equation of state simulations with the suffixes I and M, respectively. With this convention, simulations P12I and P12M refer to the P12 setup with the ideal and tabulated  equations of state, respectively; similarly, simulations N16I and N16M are the N16 simulation with the ideal and tabulated equation of state, respectively. For the N16 simulations, we also include the suffix h if the point mass particles have a softening length of $h_\text{soft}=$0.5\,\rsun. The initial parameters of all of these simulations are listed in Table~\ref{table:simulations}. Our N16M simulation will be compared to the  models of \citet{ivanova2016common} and \citet{nandez2016common}, which they called BF36 and 1.8G0.32C0.36D, respectively). We will adopt their naming convention, and refer to their model as BF36. 

\begin{table*}
\centering
\begin{tabularx}{\linewidth}{Y >{\hsize=1.35\hsize}Y *{7}{>{\hsize=0.95\hsize}Y}}
\hline
Model & $n_\text{part}$ & $R_\text{1}$ & $M_\text{1}$ & $M_\text{1,c}$ & $M_\text{gas}$ & $M_\text{2}$ & $a_\text{init}$ & EoS \\
& & (\rsun) & (\msun) & (\msun) & (\msun) & (\msun) & (\rsun) &  \\
\hline
P12I & $1.4 \times 10^{5}$ & 81 & 0.88 & 0.39 & 0.49 & 0.6 & 100 & Ideal \\
P12M & $1.4 \times 10^{5}$ & 88 & 0.88 & 0.39 & 0.49 & 0.6 & 100 & Tabulated \\
N16I & $1.4 \times 10^{5}$ & 16 & 1.80 & 0.320 & 1.479 & 0.36 & 31.37 & Ideal \\
N16M & $1.4 \times 10^{5}$ & 17 & 1.80 & 0.320 & 1.479 & 0.36 & 31.37 & Tabulated \\
N16Ih & $1.4 \times 10^{5}$ & 15 & 1.80 & 0.313 & 1.487 & 0.36 & 31.37 & Ideal \\
N16Mh & $1.4 \times 10^{5}$ & 16 & 1.80 & 0.313 & 1.487 & 0.36 & 31.37 & Tabulated \\

\hline
\end{tabularx}
\caption{Initial conditions for our simulations. The first three characters in the Model column denote which star was used (P12 or N16), the fourth character denotes which equation of state (I for ideal gas and M for the tabulated equation of state from \textsc{mesa}) and the suffix h means that the point mass particles had softening lengths of 0.5\,\rsun, instead of 1\,\rsun, $n_\text{part}$ is the number of particles in the simulation, $R_\text{1}$ is the radius of the star in the simulation after stabilisation in the computational domain, $M_\text{1}$ is the mass of the primary star, $M_\text{1,c}$ is the primary's core mass, $M_\text{gas}$ is the total gas mass in the simulation, $M_\text{2}$ is the mass of the companion, $a_\text{init}$ is the initial orbital separation and EoS shows which equation of state was used.}
\label{table:simulations}
\end{table*}

\subsection{The tabulated equation of state}
\label{ssec:eos}

\begin{figure*}
    \centering
    \subfloat[Pressure \label{fig:mesa_pressure}]{\includegraphics[width=0.49\linewidth]{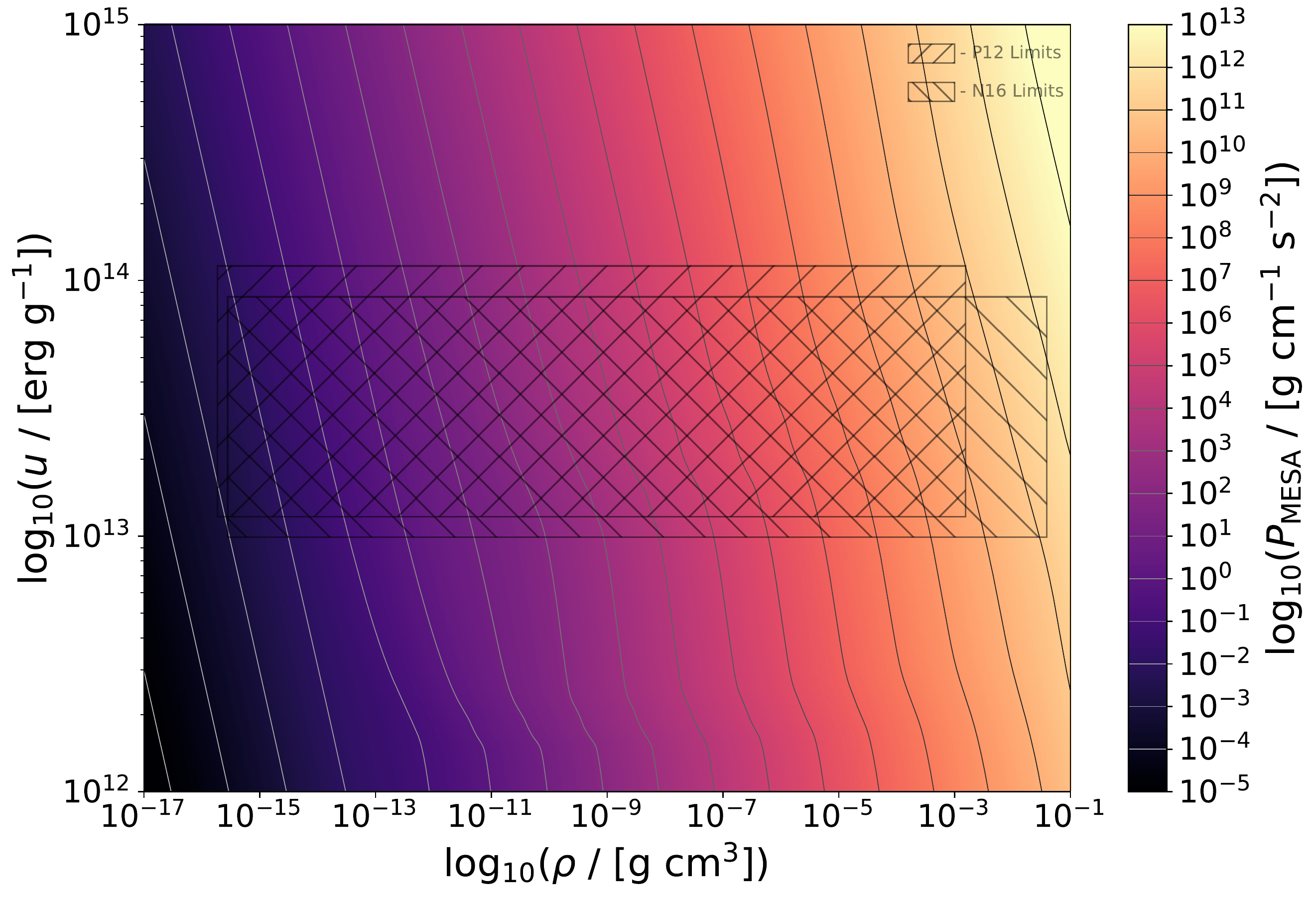}}
    \subfloat[$\Gamma_1 = \dv{\ln P}{\ln \rho}|_S$ \label{fig:mesa_gamma1}]{\includegraphics[width=0.49\linewidth]{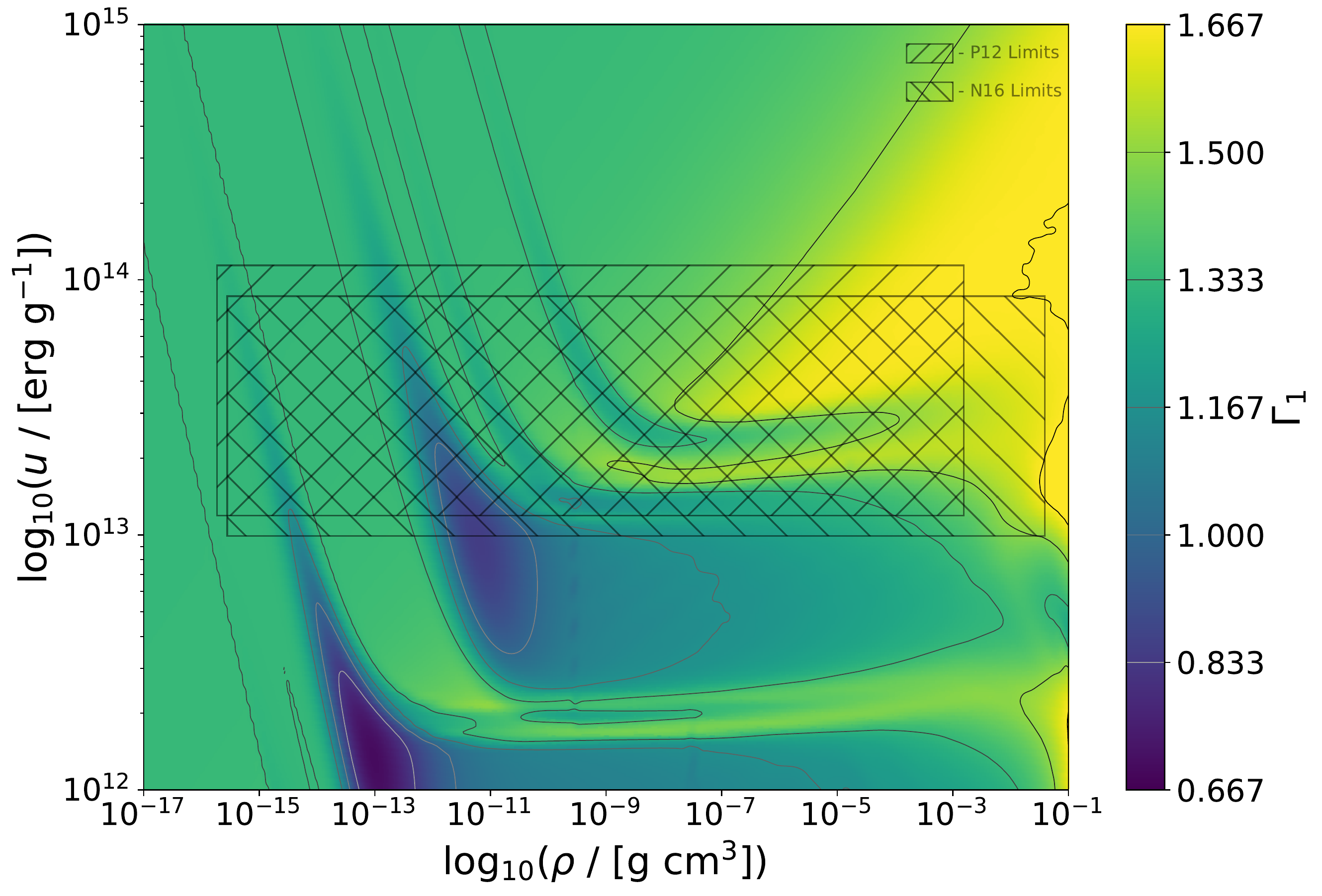}} \hfill
    \subfloat[Temperature \label{fig:mesa_temp}]{\includegraphics[width=0.49\linewidth]{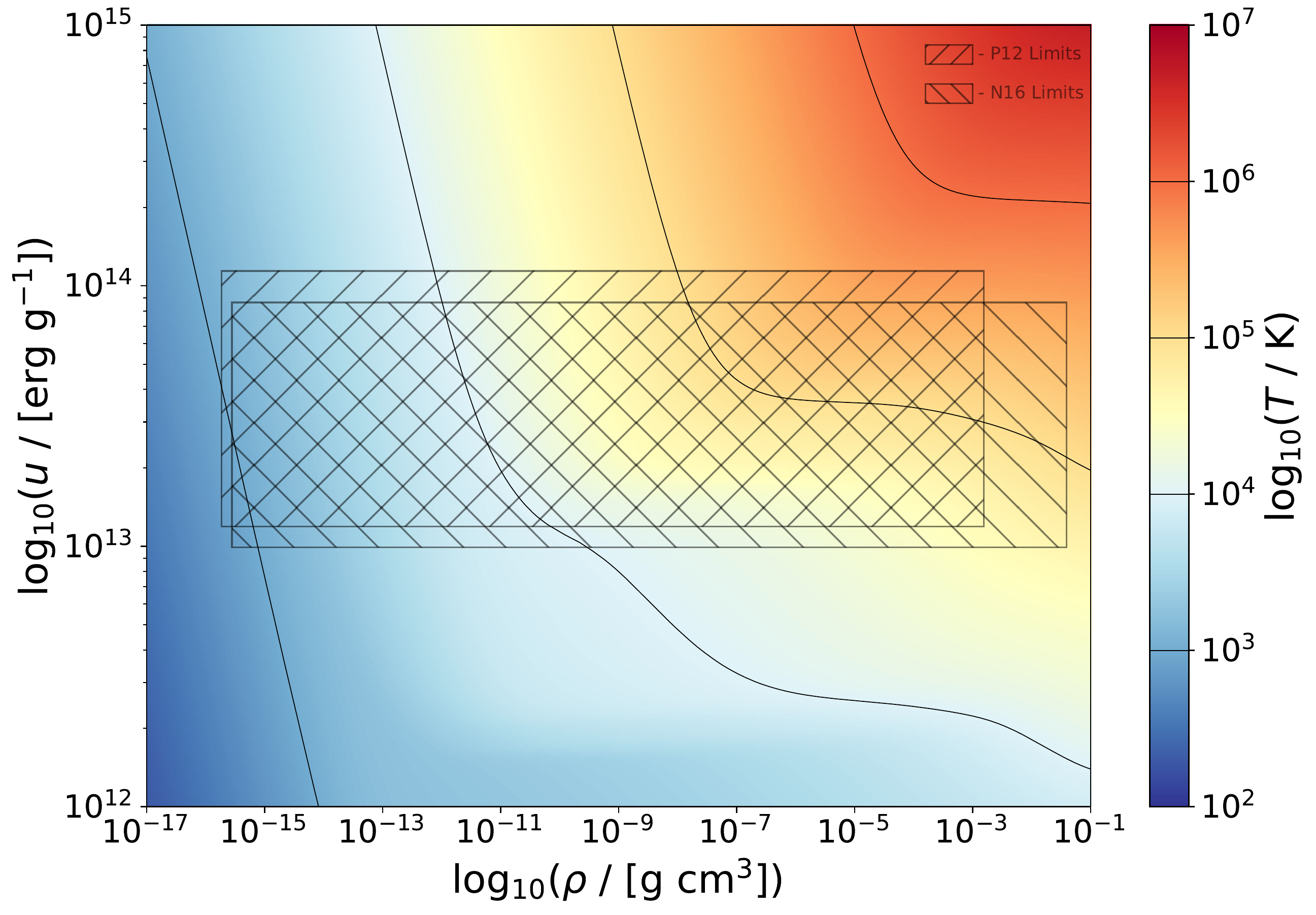}}
    \subfloat[Ratio of \textsc{mesa} gas pressure to ideal equation of state pressure \label{fig:mesa_pratio}]{\includegraphics[width=0.49\linewidth]{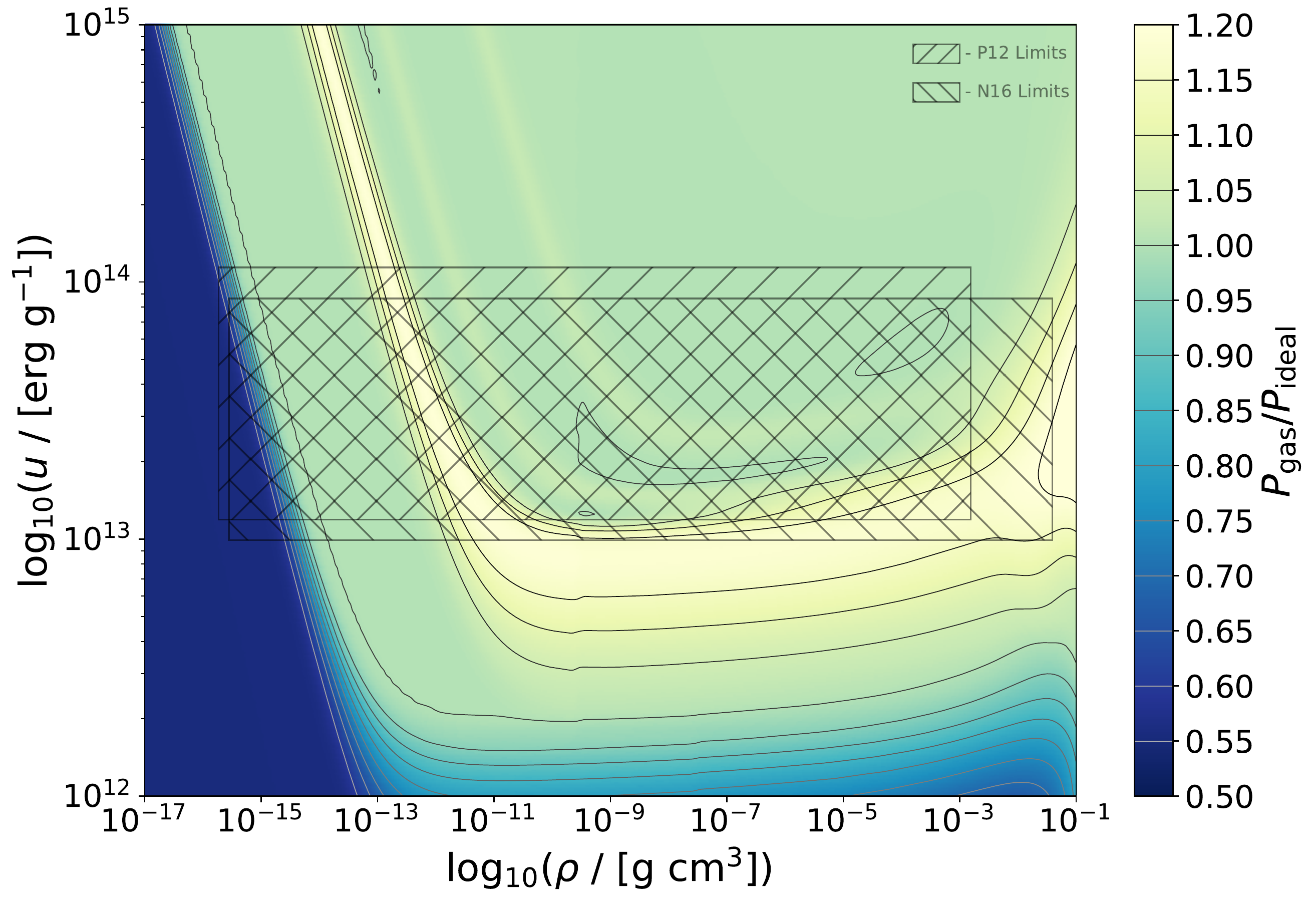}}
    \caption{Various quantities as functions of specific internal energy and density using the tabulated equation of state. The two rectangles bound all values of density and specific internal energy encountered in our simulation domains.}
    \label{fig:mesa_eos}
\end{figure*}

Our simulations employ two distinct equations of state. The first is the standard ideal gas equation of state, written in the form, \begin{equation}
P = \rho (\gamma - 1) u,
\end{equation}
where $P$ is gas pressure, $\rho$ is the gas density, $\gamma = \frac{5}{3}$ is the adiabatic index and $u$ is the specific internal energy.

The second equation of state is adapted from the equation of state used in the 1D stellar evolution code \textsc{mesa} \citep{paxton2011modules}. The \textsc{mesa} equation of state is constructed from several other equations of state. In the regions where ionisation is important, the OPAL and SCVH equation of state tables are used \citep[see section 4.2 of][]{paxton2011modules}. The OPAL equation of state \citep{rogers1996opal, rogers2002updated} is obtained by an ``activity-expansion''\footnote{Describing this method is outside the scope of this paper, though the reader is referred to \citet{rogers1994equation}, and references therein.} of the grand canonical ensemble, which includes the effects of ionisation and dissociation.

The OPAL equation of state has been used in the common envelope simulations of \citet{ohlmann2016hydrodynamic}. The SCVH equation of state \citep{saumon1995scvh} is constructed for hydrogen-helium mixtures and includes effects of temperature and pressure on ionisation and dissociation. It is intended for use in very low mass stars, as well as brown dwarfs and giant planets.

The two sets of tables overlap in a central region, within which a smooth transition is constructed between the OPAL and SCVH tables \citep[for a visualisation of this region, see Figure 1 of][]{paxton2011modules}. These tables together cover a region defined by $-17.2 \leq \log W \leq -2.9$ and $2.1 \leq \log T \leq 8.2$, where $T$ is temperature and $\log W \equiv \log P_\text{gas} - 4\log T$ is a variable introduced to allow the tables to save space by making them rectangular, as this relation describes the approximate $P_\text{gas}$--$T$ relationship of many stars. Outside the regions covered by these tables, the HELM and PC equations of state are utilised, both of which are constructed with the assumption of complete ionisation. The \textsc{mesa} equation of state tables accept $\rho$ and $T$, or $P$ and $T$, as inputs, returning the internal energy, $u$, and many other quantities, such as the entropy $S$ and $\Gamma_1 = (\partial \ln{P} / \partial \ln{\rho})_S$.

However, \textsc{phantom}, like many hydrodynamic codes, evolves the specific internal energy of the gas, using the equation of state to determine the pressure. For this reason, we constructed tables using the \textsc{mesa} \textsc{eos} module \citep{paxton2011modules}, which would accept $\rho$ and $u$ from \textsc{phantom} and return the pressure and temperature. Aside from the aforementioned pairs of input to the tables ($\rho$ and $T$ or $P$ and $T$), the \textsc{mesa} \textsc{eos} module contains a subroutine which accepts $u_\text{in}$ and $\rho_\text{in}$ to query the tables. As the tables are constructed in a fashion that has $u$ as an output, this subroutine also requires an initial guess for the temperature $T_\text{guess}$, which is used as a starting point for Newton-Raphson iterations. The tables are queried with $T_\text{guess}$ and $\rho_\text{in}$, returning a value of the specific internal energy $u_\text{out}$, which is compared to $u_\text{in}$. Also returned is the gradient of the specific internal energy with respect to temperature, at constant density. This information is enough to iterate the temperature until $u_\text{out}$ returned by the calls to the tables matches the input $u_\text{in}$. If no solution is found within the maximum number of iterations, then upper and lower bounds for the temperature are utilised in the bisection method of root-finding, which is often slower, but guarantees a root is found. When a value of $T$ is found such that $u_\text{in}$ matches $u_\text{out}$, then $P$, $\Gamma_1$ and the other equation of state values can also be returned from the tables.

The new tables cover a region defined by $-10 \leq \log V \leq 12$ and $10 \leq \log u \leq 17$, where $u$ is specific internal energy and $\log V \equiv \log \rho - 0.7\log u + 20$ is again a new variable introduced to make use of the approximate relationship between $\rho$ and $u$ in stars, saving space in the tables. The tables are produced for hydrogen mass fractions of $X =$ 0, 0.2, 0.4, 0.6 and 0.8, each with metals mass fractions of $Z =$ 0, 0.02 and 0.04. We wrote a set of routines to query these tables from within \textsc{phantom}. We plot the pressure, $\Gamma_1$ and temperature returned by our tabulated equation of state tables in Figs~\ref{fig:mesa_pressure}, \ref{fig:mesa_gamma1} and \ref{fig:mesa_temp}, respectively, when the tables are queried through \textsc{phantom}. In Fig.~\ref{fig:mesa_pratio}, we show a comparison of the gas pressure returned by our tabulated equation of state to the equivalent pressure from the ideal gas equation of state. The differences here lie primarily in the fact that the ideal gas equation of state has constant mean molecular weight. The ranges of $\rho$ and $u$ over the course of the entire P12M and N16M simulations are overlaid as rectangles, showing that the limits fall well within the boundaries of the tables. We also plot contours to give a better understanding of how the returned values change with $\rho$ and $u$.

\section{Results}
\label{sec:results}

\begin{table*}
\centering
\begin{minipage}{0.75\linewidth}
\begin{tabularx}{\linewidth}{Y Y Y Y Y Y Y Y}
\hline
Model &  $E_\text{pot}$ &  $E_\text{th+r}$ &  $E_\text{rec}$ & $E_\text{orb}$ & $E_\text{bin}$ & $E_\text{tot}$ & $J_\text{tot}$ \\
 & ($10^{46}$\,erg) & ($10^{46}$\,erg) & ($10^{46}$\,erg) & ($10^{46}$\,erg) & ($10^{46}$\,erg) & ($10^{46}$\,erg) & ($10^{52}$\,erg\,s) \\
\hline
P12I  &  $-5.12$ &  2.43 &     $-$ &  $-1.82$ & $-2.68$ &  $-3.67$ &  2.61 \\
P12M  &  $-5.25$ &  2.53 &  1.27 &  $-1.80$ & $-2.71$ &  $-2.28$ &  2.61 \\
N16I  &  $-85.4$ &  42.0 &     $-$ &  $-3.94$ & $-43.3$ &  $-47.3$ &  1.49 \\
N16M  &  $-87.0$ &  43.3 &  4.35 &  $-4.07$ & $-43.7$ &  $-42.7$ &  1.49 \\
N16Ih &  $-86.8$ &  43.0 &     $-$ &  $-3.91$ & $-43.8$ &  $-47.7$ &  1.49 \\
N16Mh &  $-88.5$ &  44.4 &  4.37 &  $-4.00$ & $-44.1$ &  $-43.1$ &  1.49 \\
BF36\textsuperscript{\textdagger}  &  $-88.1$ &  44.0 &  4.68 &  $-3.91$ & $-44.2$ &  $-43.4$ &  1.49 \\
\hline
\multicolumn{6}{l}{\textsuperscript{\textdagger}\footnotesize{Simulation presented in \citet{ivanova2016common}.}}
\end{tabularx}
\caption{Values of energies and angular momenta at the beginning of the simulations. The model column is the same as in Table~\ref{table:simulations}. $E_\text{pot}$, $E_\text{th+r} = E_\text{therm} + E_\text{rad}$ and $E_\text{rec}$ are the gravitational potential energy, the sum of thermal and radiation energies and the recombination energy of the red giant's envelope, respectively. $E_\text{orb}$, $E_\text{tot}$ and $J_\text{tot}$ are the orbital energy, the total energy and the total angular momentum of the binary system, respectively.}
\label{table:energies}
\end{minipage}
\end{table*}

\begin{table*}
\begin{tabularx}{\linewidth}{*{5}{>{\hsize=0.94\hsize}Y}*{3}{>{\hsize=1.3\hsize}Y}*{2}{>{\hsize=0.7\hsize}Y}}
\hline
Model & $M_\text{b,PK}$ & $M_\text{b,PKU}$ & $M_\text{b,PKT}$ & $M_\text{b,PKTR}$ & $E_\text{u,tot}$ & $J_\text{u,tot}$ & $E_\text{orb}$ & $t$ & $a$ \\
 & (\%) & (\%) & (\%) & (\%) & ($10^{46}$\,erg) & ($10^{52}$\,erg\,s) & ($10^{46}$\,erg) & (d) & (\rsun) \\
\hline
\multicolumn{10}{c}{Values at 50\,d after the end of their dynamical inspiral}\\
\hline
P12I  &  84 &  84 &     $-$ &     $-$ &  0.15 &  0.54 &  $-3.76$ &  300 &  18.8 \\ 
P12M  &  69 &   6.1 &  65 &  61 &  0.52 &  0.99 &  $-3.31$ &  359 &  18.6 \\ 
N16I  &  93 &  93 &     $-$ &     $-$ &  1.11 &  0.42 &  $-59.1$ &  157 &  0.60 \\ 
N16M  &  92 &  76 &  90 &  90 &  1.76 &  0.51 &  $-59.5$ &  158 &  0.58 \\ 
N16Ih &  87 &  81 &     $-$ &     $-$ &  1.49 &  0.59 &  $-54.3$ &  157 &  0.47 \\ 
N16Mh &  55 &  26 &  50 &  50 &  7.19 &  0.91 &  $-59.7$ &  138 &  0.43 \\ 
\hline
\multicolumn{10}{c}{Values at the end of the simulations}\\
\hline
P12I  &   41 &  37 &     $-$ &     $-$ &  0.39 &  1.51 &  $-3.68$ &  1843 &  12.5 \\
P12M  &   12 &   2.0 &   6.1 &   2.0 &  1.24 &  2.05 &  $-3.19$ &  1843 &  14.5 \\
N16I  &   92 &  93 &     $-$ &     $-$ &  1.09 &  0.41 &  $-52.8$ &   922 &  0.51 \\
N16M  &   63 &   1.4 &  45 &  43 &  2.72 &  1.24 &  $-50.0$ &   922 &  0.49 \\
N16Ih &   43 &  37 &     $-$ &     $-$ &  2.09 &  1.79 &  $-52.4$&   553 &  0.37 \\
N16Mh &    0.2 &   0.0 &  <0.1 &  <0.1 &  15.8 &  1.50 &  $-61.1$ &   553 &  0.44 \\
\hline
\end{tabularx}
\caption{We present the bound mass in the envelope of the primary at two times during the simulations and for different definitions of unbound mass. If the energy of the particle is negative, it is marked as bound to the system. The subscripts `P', `K', `U', `T' and `R' refer to potential, kinetic, total internal, thermal and radiation energies, respectively, which are summed to give the particle energy relevant to unbinding. $E_\text{u,tot}$ and $J_\text{u,tot}$ are the total energy and angular momentum of the unbound material, respectively, when summing the kinetic, potential and thermal energies to determine unbound material. $E_\text{orb}$ is the orbital energy of the binary system. Columns $t$ and $a$ refer to the time from the beginning of the simulation and orbital separation at which these measurements were taken. All masses are given as percentages of the total gas mass in their simulations (which are given in Table~\ref{table:simulations}).}
\label{table:bound}
\end{table*}

Here we analyse the behaviour of the simulations as they undergo the fast inspiral and the subsequent phase of slow inspiral. We examine the decrease in bound mass in each simulation, and particularly note the differences between simulations with different equations of state.

\begin{figure*}
	\includegraphics[width=0.75\linewidth]{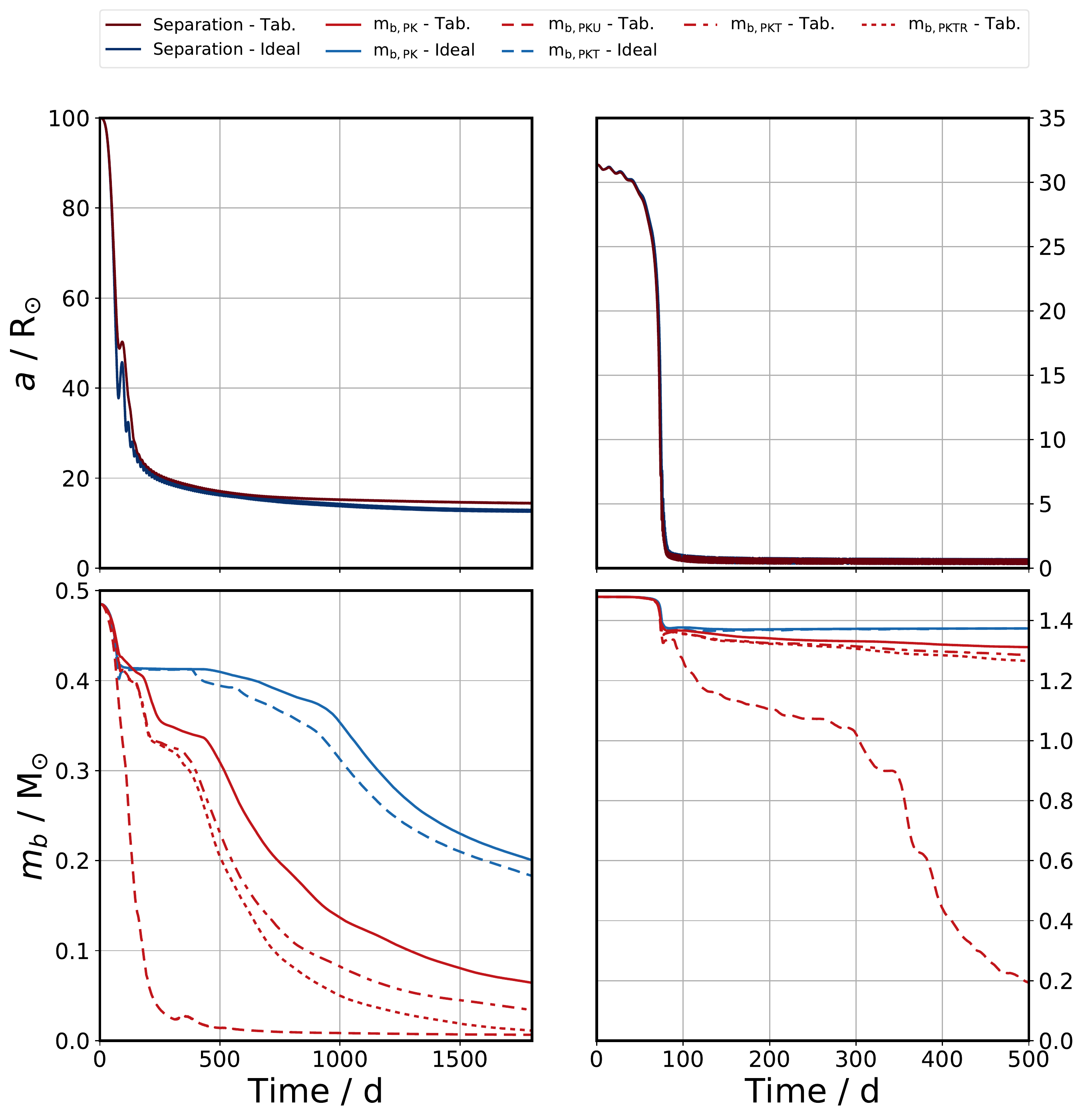}
    \caption{Top row: Orbital separation of the two point mass particles. Bottom row: Evolution of the bound mass in each simulation. In the legend, the subscripts refer to the criterion used to determine if a particle was unbound. The subscripts P, K, U, T and R refer to potential, kinetic, total internal, thermal and radiation energies, respectively. That is, for the subscript PK, the mass is unbound if the sum of its potential and kinetic energies are positive. Left column: P12 simulations. Right column: N16 simulations. The blue lines refer to simulations with the ideal equation of state, and the red lines to simulations with the tabulated equation of state. In the case of the ideal gas, the thermal and internal energies are the same (that is, the subscripts U and T refer to the same energy).}
    \label{fig:sep_bound}
\end{figure*}

\subsection{Energy and angular momentum}

Energy conservation for the P12I and P12M simulations is approximately at the 0.1~percent level, while for the N16I and N16M simulations is at about the 2~percent level. The total angular momentum in all of the simulations is conserved to about 0.1~percent. We list key energy and angular momentum values, in Table~\ref{table:energies} and compare them to the values given by \citet{ivanova2016common}. The total energy budget of our N16M simulation differs from the total energy of BF36 by less than 1~percent, and the total angular momentum budget differs by less than 0.1~percent. 

We define how we have calculated the various energy components within this work. Within \textsc{phantom}, and most other SPH codes, each particle $i$ has a mass $m_i$, a density $\rho_i$, a specific internal energy $u_i$ and a specific gravitational potential energy $\phi_i$, along with position and velocity vectors, $\vec{x}_i$ and $\vec{v}_i$. The total gravitational potential energy is calculated as \begin{equation}
    E_\text{pot} = \sum_i m_i \phi_i
\end{equation}
which sums the potential energies of each particle in the simulation. We determine the thermal and radiation energies in the tabulated equation of state simulations as follows:
\begin{align}
    E_\text{therm} =& \sum_i m_i \frac{3}{2} \frac{k_B T_i}{\mu_i m_H}\\
    E_\text{rad} =& \sum_i m_i \frac{a T_i^4}{\rho_i}
\end{align}
where $k_B \approx 1.38 \times 10^{-16}$\,erg\,K$^{-1}$ is the Boltzmann's constant, $T_i$ is the temperature of the particle, $\mu_i$ is the mean molecular weight of the particle, and $m_H \approx 1.67 \times 10^{-24}$\,g is the mass of a hydrogen atom.

The total recombination energy reservoir available in the star can be calculated by performing the following sum:
\begin{equation}
    E_\text{rec} = \sum_{i} E_\text{i;rec}, \label{eq:erec}
\end{equation}
where the recombination energy of a single particle $E_\text{i;rec}$, is given by:
\begin{equation}
    E_\text{i;rec} = m_i N_A \left(X x_{\text{HII}} \chi^0_\text{H} + \frac{Y}{4} \left[\chi^0_\text{He} \left(x_{\text{HeII}} + x_{\text{HeIII}}\right) + \chi^1_\text{He} x_{\text{HeIII}}\right]\right), \label{eq:ereci}
\end{equation}
where $m_i$ is the mass of the particle, $N_A = 6.02 \times 10^{23}$~mol$^{-1}$ is Avogadro's number, $X$ and $Y$ are the hydrogen and helium mass fractions of the SPH particle, $x_{\text{HII}}$ is the fraction of ionised hydrogen, $x_{\text{HeII}}$ and $x_{\text{HeIII}}$ are the fractions of singly and doubly ionised helium, respectively. $\chi^0_\text{H} = 13.6$\,eV is the ionisation energy of neutral hydrogen and $\chi^0_\text{He} = 26.4$\,eV and $\chi^1_\text{He} = 54.4$\,eV are the ionisation energies of neutral and singly ionised helium, respectively. Here we neglect the contributions of elements heavier than helium, as these contributions will be small. The quantities $m_i N_A X$ and $\frac{1}{4}m_i N_A Y$ estimate the number of atoms of hydrogen and helium that are present within one SPH particle.

The orbital energy is determined using the same method as \citet{nandez2015recombination}. That is, we calculate:
\begin{equation}
    E_\text{orb} = \frac{1}{2} \left( \mu |\vec{v}_{1} - \vec{v}_{2}|^2 + \sum_i m_i \phi_i - \sum_j m_j \phi_j^{\text{RL}_1}\textbf - \sum_k m_k \phi_k^{\text{RL}_2}\right)
\label{eq:e_orb}
\end{equation}
where $\mu = M_1 M_2/(M_1 + M_2)$ is the reduced mass of the binary system and $\vec{v}_{1}$ and $\vec{v}_{2}$ are the velocity vectors of the two core particles. The second term in Eq.~\ref{eq:e_orb} expresses the total gravitational potential energy of the system by summing over all particles $i$, while the third and fourth terms subtract the components relating to the gravitational attraction of particles within both Roche lobes (summing over the particles in $RL_1$, index $j$, and the particles in $RL_2$, with index $k$). The remainder is the portion of the gravitational potential energy which contributes to the orbital energy.

We calculate the binding energy of our star by summing the thermal, radiation and gravitational potential energies of the star in isolation:
\begin{equation}
    E_\text{bin} = E_\text{therm} + E_\text{rad} + E_\text{pot}.
\end{equation}
Note that we do not include the recombination energy. In the ideal gas equation of state simulations, we calculate the binding energy without the radiation pressure term, as our simulations do not include it.

Both the gravitational potential energy and the sum of thermal and radiation energies of our N16M primary red giant differ by about 1~percent from the primary star used in BF36. The largest disparity comes from the approximately 7~percent difference in the recombination energy term. This can potentially be explained by the fact that we assigned the same hydrogen and helium mass fractions to all of our SPH particles, while \citet{ivanova2016common} used the abundances of their input \textsc{mesa} profile to assign different H and He mass fractions to each of their particles.

Finally, the orbital energy in our simulation is larger than for the BF36 simulation of \citet{nandez2016common} by approximately 4~percent. This is likely due to the fact that our star stabilises at a slightly larger radius. In conclusion, we can state that the initial conditions of our N16M simulation are sufficiently similar to those of the BF36 model to enable the comparison of key quantities.

\subsection{The orbital separation evolution}

The orbital evolution for our simulations can be seen in the top panels of Fig.~\ref{fig:sep_bound}, where the left and right columns are for the P12 and N16 simulations, respectively. The blue lines in both columns correspond to simulations utilising the ideal equation of state, while the red lines pertain to the tabulated equation of state simulations. 

By examining Fig.~\ref{fig:sep_bound}, we see that, in both simulation setups, the orbital evolution is relatively unaffected by the choice of equation of state. Table~\ref{table:bound} gives values for the orbital separations just after the inspiral and at the end of the simulations. In the P12I and P12M simulations, the orbital separations 50\,d after the end of the inspiral are 18.8\,\rsun\ and 18.6\,\rsun, respectively, while for the N16I and N16M simulations, the separations are 0.60\,\rsun\ and 0.58\,\rsun, respectively. Indeed, by the end of the simulations, N16I and N16M simulations have almost identical orbital separations of 0.51\,\rsun\ and 0.49\,\rsun, respectively, while the P12I final separation is somewhat smaller than the P12M simulation (12.5\,\rsun\ and 14.5\,\rsun, respectively). The slight variations could easily be due to the fact that the initial stellar structures have stabilised at slightly different radii with the different equations of state. {\it This suggests that the amount of orbital energy deposited into the envelope during the dynamic inspiral is independent of the equation of state.} We conclude that any extra unbinding that occurs with the tabulated equation of state is not  caused by the input of extra orbital energy.

\subsection{The evolution of the unbound mass}
\label{ssec:bound}

The strictest criterion for determining whether or not mass is unbound from the system is to use the threshold $E_\text{kin} + E_\text{pot} > 0$. Assuming that unbound SPH gas particles are not trapped below a layer of bound particles to which they can transfer kinetic energy, this means that the unbound material will never return to interact with the compact binary system. Some portion of the internal energy of the gas particle can also be included in the energy balance, which results in a greater amount of unbound mass (see the comparison carried out by \citealt{iaconi2017effect}).

For the simulations using an ideal gas equation of state, the internal energy of the gas corresponds exactly to its thermal energy, that is, the disordered kinetic energy of the gas particles (rather than the bulk motion of the gas). Therefore, the total gas energy, comprising kinetic, potential and thermal energies, can be used to determine whether the gas is bound or unbound. However, in those simulations that use the tabulated equation of state, the internal energy of the gas does not only include the thermal energy, but also the energy associated with radiation and that associated with the possible recombination of hydrogen, helium and metals. While the gas remains ionised, the recombination energy is \textit{latent}, and does not affect the system's dynamics. Therefore,  including  recombination energy in the internal energy, overestimates the amount of unbound gas. On the other hand, when expanding envelope gas recombines, the latent energy is released and immediately thermalised, therefore contributing to the thermal energy of the gas. For fully recombined gas, using the internal energy of the gas is therefore equivalent to using the thermal and radiation energies in ideal gas simulations.

Because of the complexities of determining whether SPH particles are unbound and the difficulties in comparing simulations using different equations of state, we calculate the amount of unbound gas using four criteria. These are \textit{(i)} the strict mechanical limit, including only kinetic and potential energies; \textit{(ii)} the sum of kinetic, potential and thermal energies (with $u_\text{therm} = 3 k_B T/2 \mu m_H$, where $k_B$ is the Boltzmann constant, $T$ is the temperature, $\mu$ is the mean molecular weight, and $m_H$ is the mass of a hydrogen atom); \textit{(iii)} the sum of kinetic, potential, thermal and radiation energies ($u_\text{therm} + a T^4/\rho$, where $a$ is the radiation constant and $\rho$ is the density); and \textit{(iv)} the sum of kinetic, potential and gas internal energy (which includes the latent recombination energy term in the case of the tabulated equation of state). For the two simulations using the ideal gas equation of state, we can only use the first two criteria (criteria \textit{(ii)} and \textit{(iv)} are actually identical, and the simulations do not include the effects of radiation pressure, hence we cannot use criterion \textit{iii}). 
 
Values of the bound masses calculated using these criteria are presented in Table~\ref{table:bound}. The ideal gas simulations consistently have similar or more mass remaining bound to the system, both 50\,d after the inspiral and at the end of the simulations. We note that, at the end of the P12M simulation, only 12~percent of the gaseous common envelope is still bound, while for the N16M simulation 63 percent is still bound (note that these two figures are upper limits because we are using the strictest criterion (i)). In contrast, the ideal gas equation of state counterpart simulations P12I and N16I have 41 and 92~percent of the envelope still bound, respectively. The difference between the two equations of state is considerably starker at the end of the simulations  (Table~\ref{table:bound}). This suggests that the majority of the unbinding that occurs during the inspiral is not influenced by recombination energy, and is instead a direct result of the injection of orbital energy, while further unbinding after the fast inspiral is aided by the release of recombination energy.

We plot the evolution of the bound mass, using all four criteria in the lower two panels of Fig.~\ref{fig:sep_bound}. We note two interesting features. First, when using the tabulated equation of state, the mass is unbound at a much greater rate than when using the ideal equation of state. This is true regardless of which criterion we use to define the bound mass. Second, as expected, the unbinding rate when using the tabulated equation of state is by far the largest if the entire internal energy budget is used in the definition. However, this implies that all gas fully recombines, and that the entire recombination energy budget is used to unbind the gas.

The P12M simulation unbinds almost the entire envelope by the time at which we end it, no matter what the definition of the bound mass (see the red lines in the bottom-left panel of Fig.~\ref{fig:sep_bound}). On the other hand, the N16M simulation only unbinds a large fraction of the envelope if we use the full internal energy expression, which includes the latent recombination energy. If we use any of the other three definitions, the unbinding is only marginally larger than when using an ideal gas equation of state. 

\citet{nandez2016common} concluded that the entire envelope is unbound. To determine whether gas is unbound, they used the sum of gas kinetic, potential and internal energies (our fourth criterion), which includes the latent recombination energy of ionised gas. As we have described above, this criterion over-estimates the fraction of unbound gas. 

Below we delve further into possible reasons why different criteria lead to such difference in the amount of unbound gas.

\subsection{Simulations with smaller softening length}
\label{ssec:movement_out_of_plane}

The softening length dictates the strength of the interaction between the gas particles and the cores, introducing a second parameter, after resolution, that may alter results. Unfortunately, the large wall clock times of these simulations have precluded a proper convergence test. However, here we compare our N16I and N16M simulations, using a softening length $h_\text{soft} = 1$\,\rsun, with simulations using $h_\text{soft} = 0.5$\,\rsun. We call the small-$h_\text{soft}$ simulations N16Ih and N16Mh (see Table~\ref{table:simulations}).

The ideal gas, smaller softening length simulation N16Ih, unbinds 6~percent more gas by 50\,d after the end of the dynamic inspirals compared to the corresponding simulation with a larger softening length (N16I) and 49 percent more by the end of the simulation. In the case of simulations using a tabulated equation of state, the small softening length simulation, N16Mh, unbinds approximately 37~percent more by 50 days after inspiral than the corresponding large softening length simulation (N16M) and 63 percent more by the end of the simulation.

We suggest two reason for this dramatic difference. Shortly after the dynamic inspiral during the N16Ih and N16Mh simulations, the binary acquires a 10\,km\,s$^{-1}$ velocity in the positive z-direction, while concomitantly a plume of gas moves in the negative z-direction (linear momentum, energy and angular momentum in these simulations are conserved to better than the 3~percent level). We attribute this to initial asymmetries in the SPH particle distribution about the orbital plane, which become exaggerated in the stronger gas-core interaction of the smaller softening length simulations. The displacement of the binary from the orbital plane and concomitant movement of some of the gas in the opposite direction reduces the gravitational potential energy of the system, leading to an increase in the unbound mass. 

A second effect is that the more rapid expansion of the gas in the N16Mh simulation compared to the N16M one leads to a release of recombination energy earlier on, leading to the faster unbinding we observe. In fact, even in  simulation N16M, with a larger softening length, the unbound mass using any of the definitions is still decreasing by the end of the simulation, implying that more, or all, of the envelope gas could be unbound if the simulation were continued for a longer time. 

We note finally an unresolved issue that was already described by \citet{reichardt2019extending}. In both the simulations with the large and small softening lengths, the gas particles directly around the sinks have SPH smoothing lengths of approximately 0.2\,\rsun. This is smaller than both the large and small tested softening lengths of the cores (1 and 0.5~\rsun, respectively), implying that the softening lengths are resolved by the gas in both cases. This said, the smaller softening length is significantly less resolved and this may have an impact on the simulations. More simulations are needed to test this effect. 

In conclusion, the smaller softening length leads to a faster interaction and earlier unbinding. However, if simulations were evolved for longer, it does appear that, independent of the softening length, recombination energy would lead to unbinding the great majority of the envelope.

\section{The cause and effect relationship between recombination energy and envelope unbinding}

To determine whether or not it is the released recombination energy that is responsible for the extra amount of unbound gas, we here determine the locations of gas unbinding and different ionisation stages of hydrogen and helium. We can then use this information to ascertain the amount of recombination energy being released, and check whether or not this released energy is spatially coincident with newly unbound gas.

\subsection{Where is gas recombining?}
\label{ssec:where_recombination}
\begin{figure*}
    \centering
    \subfloat[Distribution of HII/H in the P12M simulation \label{fig:p12_HII}]{\includegraphics[width=0.49\linewidth]{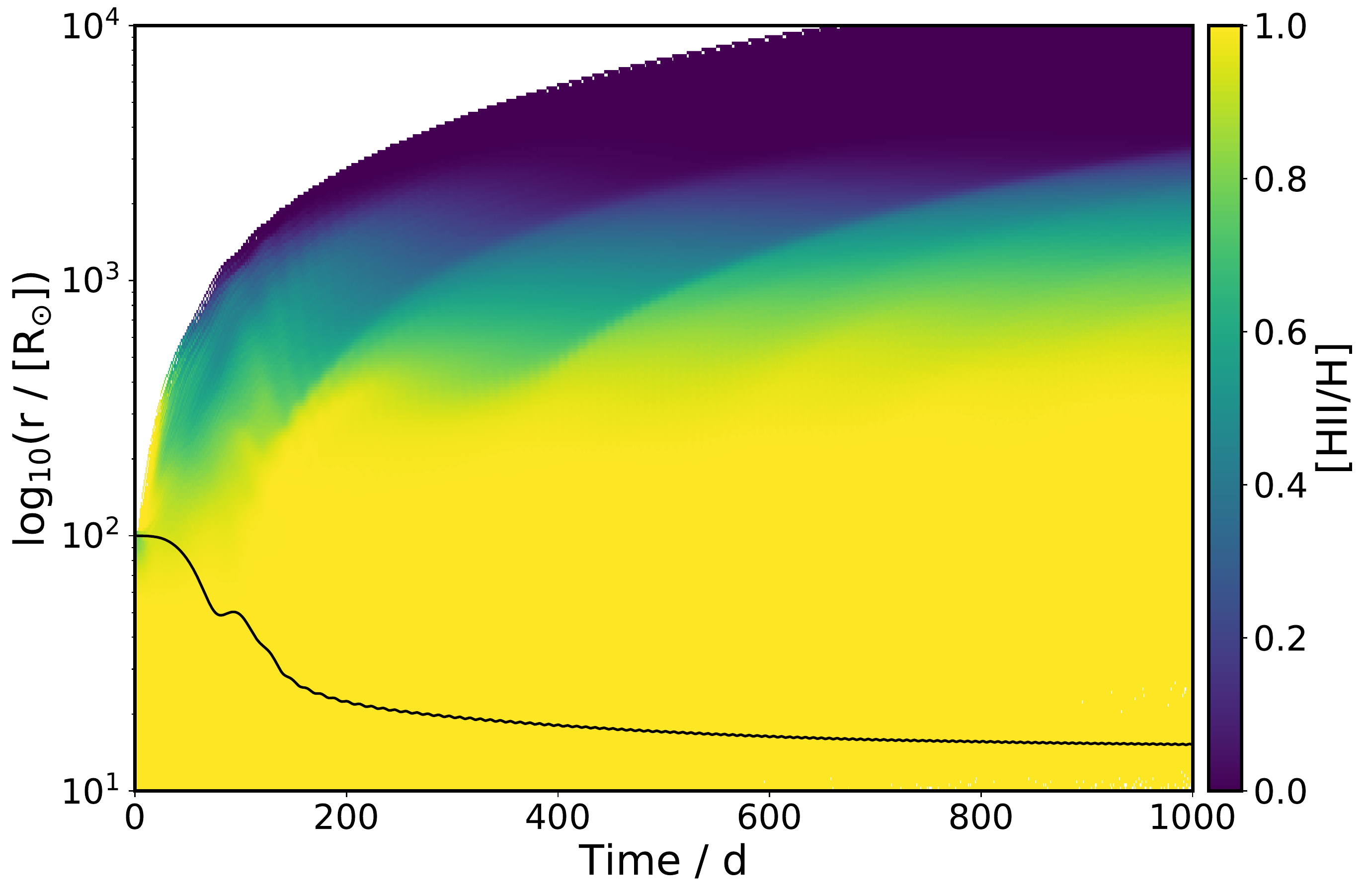}}
    \subfloat[Distribution of HII/H in the N16M simulation \label{fig:n16_HII}]{\includegraphics[width=0.49\linewidth]{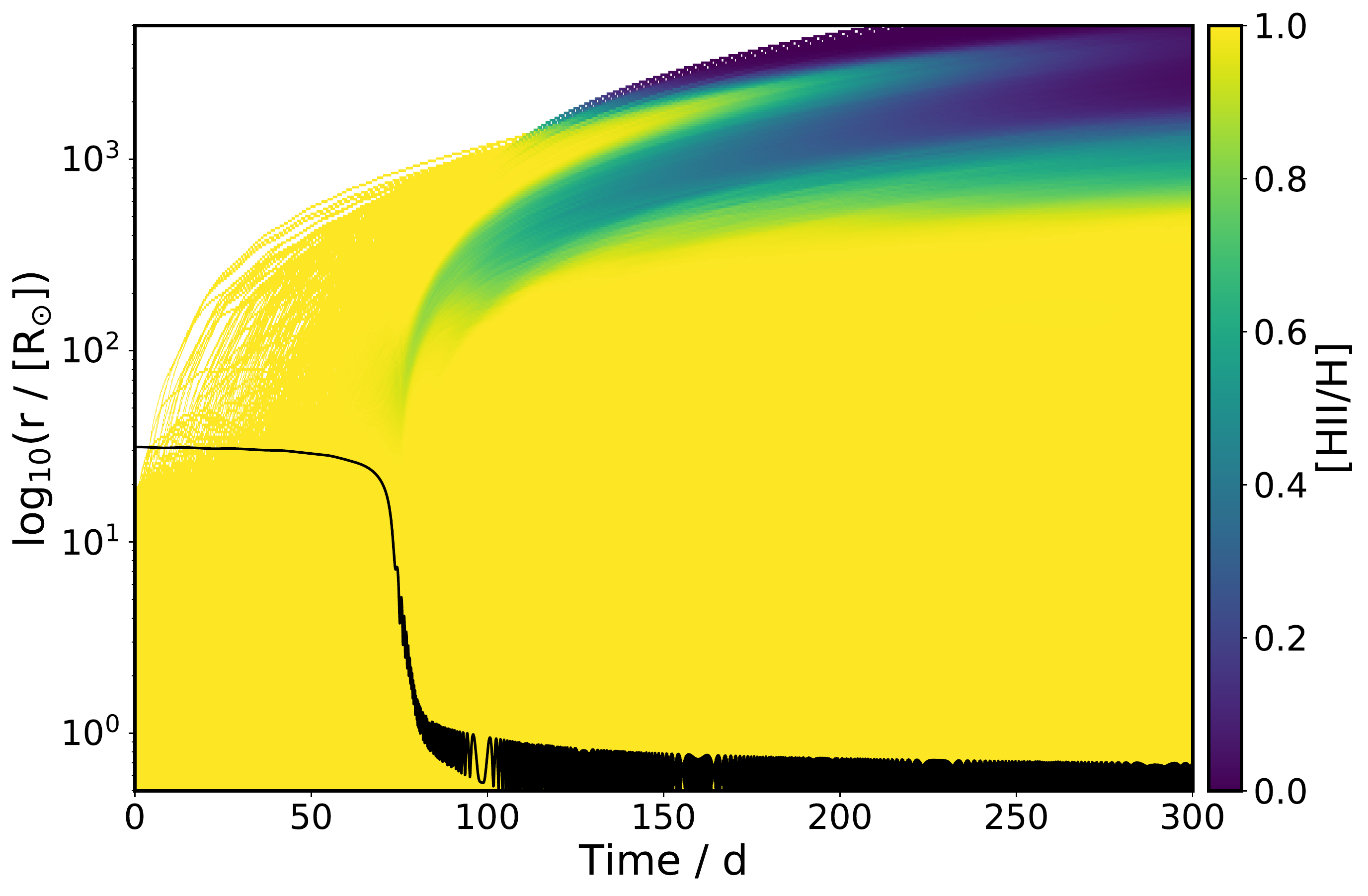}}\\
    \subfloat[Distribution of HeII/H in the P12M simulation \label{fig:p12_HeII}]{\includegraphics[width=0.49\linewidth]{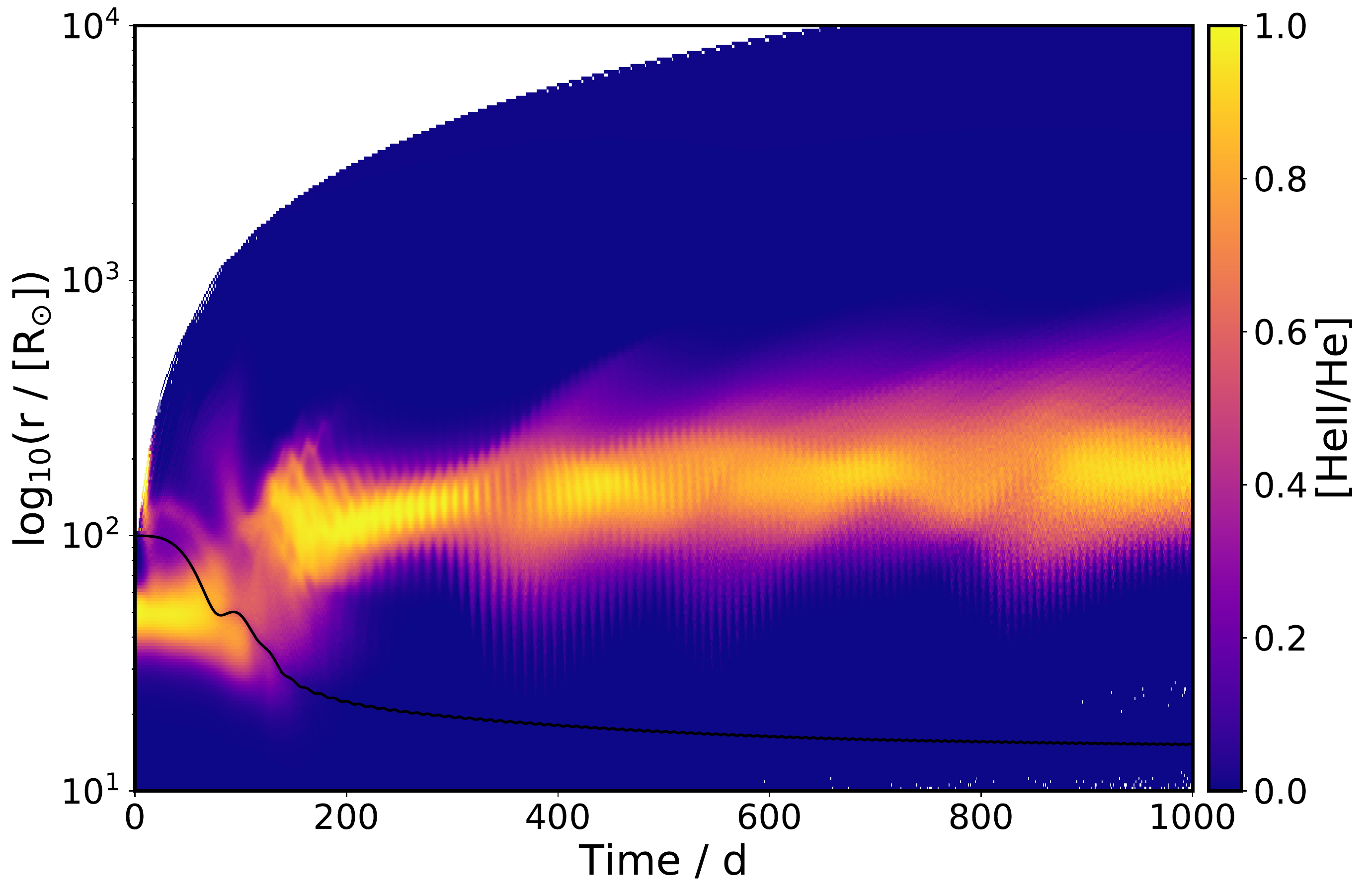}}
    \subfloat[Distribution of HeII/H in the N16M simulation \label{fig:n16_HeII}]{\includegraphics[width=0.49\linewidth]{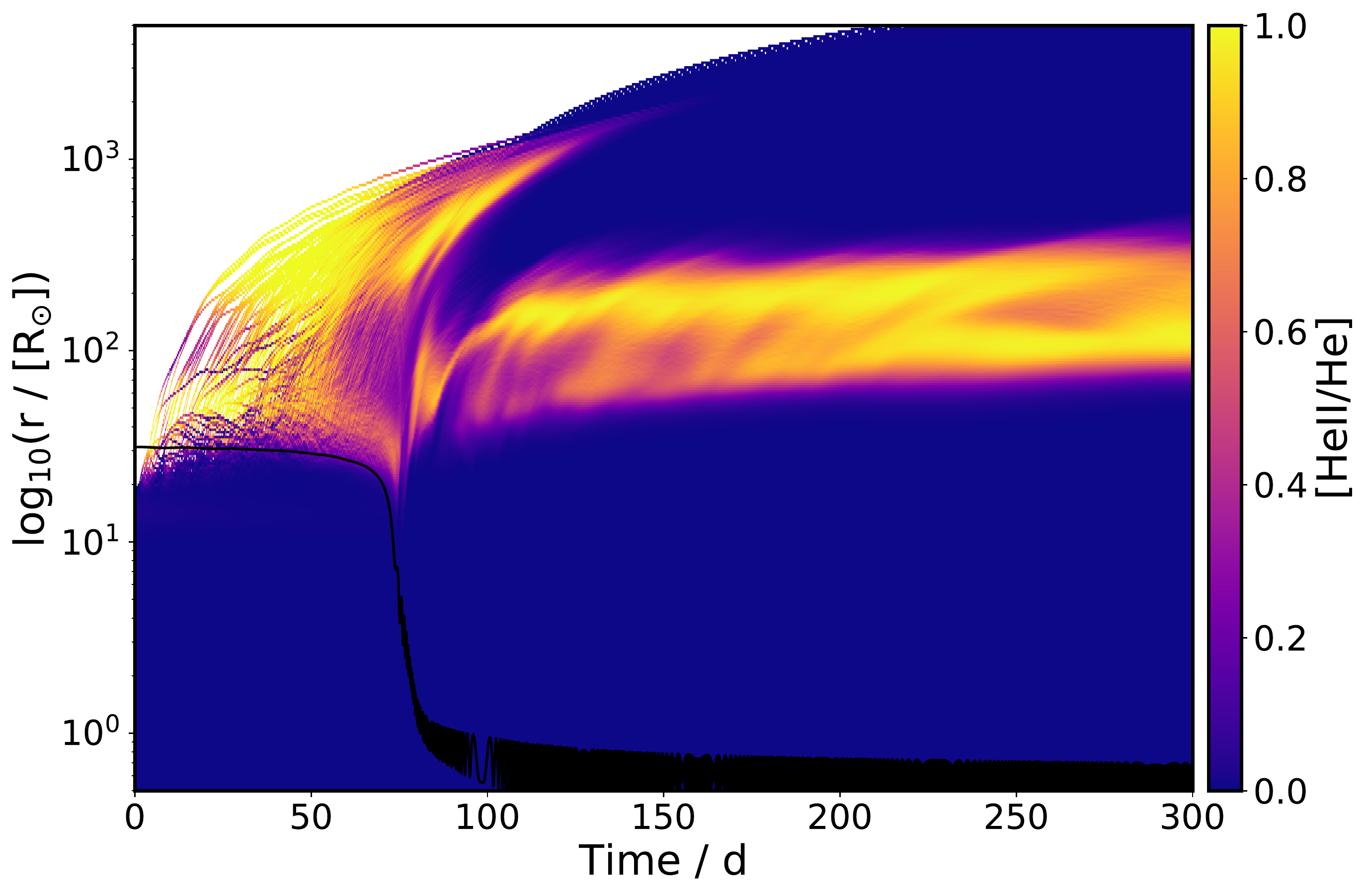}}
    \caption{The ionisation stages of H and He in the P12M (left column) and N16M simulations (right column) as a function of radius and time. Top row: Radially averaged amount of HII as a fraction of the total amount of H. Bottom row: Radially averaged amount of HeII as a fraction of the total amount of He. The radii are centred on the core of the primary star. The black line depicts the distance of the companion from the core of the primary.}
    \label{fig:ions}
\end{figure*}

 To determine the ionisation state of hydrogen and helium as a function of envelope depth and of time we use the Saha equation: 
\begin{equation}
\frac{n_{i+1} n_e}{n_i} = \frac{2}{\lambda} \frac{g_{i+1}}{g_i} \exp\left[-\frac{\epsilon_{i+1} - \epsilon_i}{k_B T} \right],
\end{equation}
where $n_i$, $g_i$ and $\epsilon_i$ are the number density, degeneracy of states and ionisation energy of ions in the $i$-th state of ionisation, respectively, $n_e$ is the number density of electrons. The parameter $\lambda \equiv \sqrt{h/(2\pi m_e k_B T)}$, where $h$ is Planck's constant, $m_e$ is the mass of an electron, $k_B$ is Boltzmann's constant and $T$ is temperature. We can track the recombination of both hydrogen and helium by simultaneously solving three Saha equations for HI, HeI and HeII. By tracking the ionisation fractions of both hydrogen and helium, we can begin to understand whether it is hydrogen or helium recombination energy that is available for unbinding the envelope.
\label{ssec:where_unbound}
\begin{figure*}
    \centering
    \includegraphics[width=0.49\linewidth]{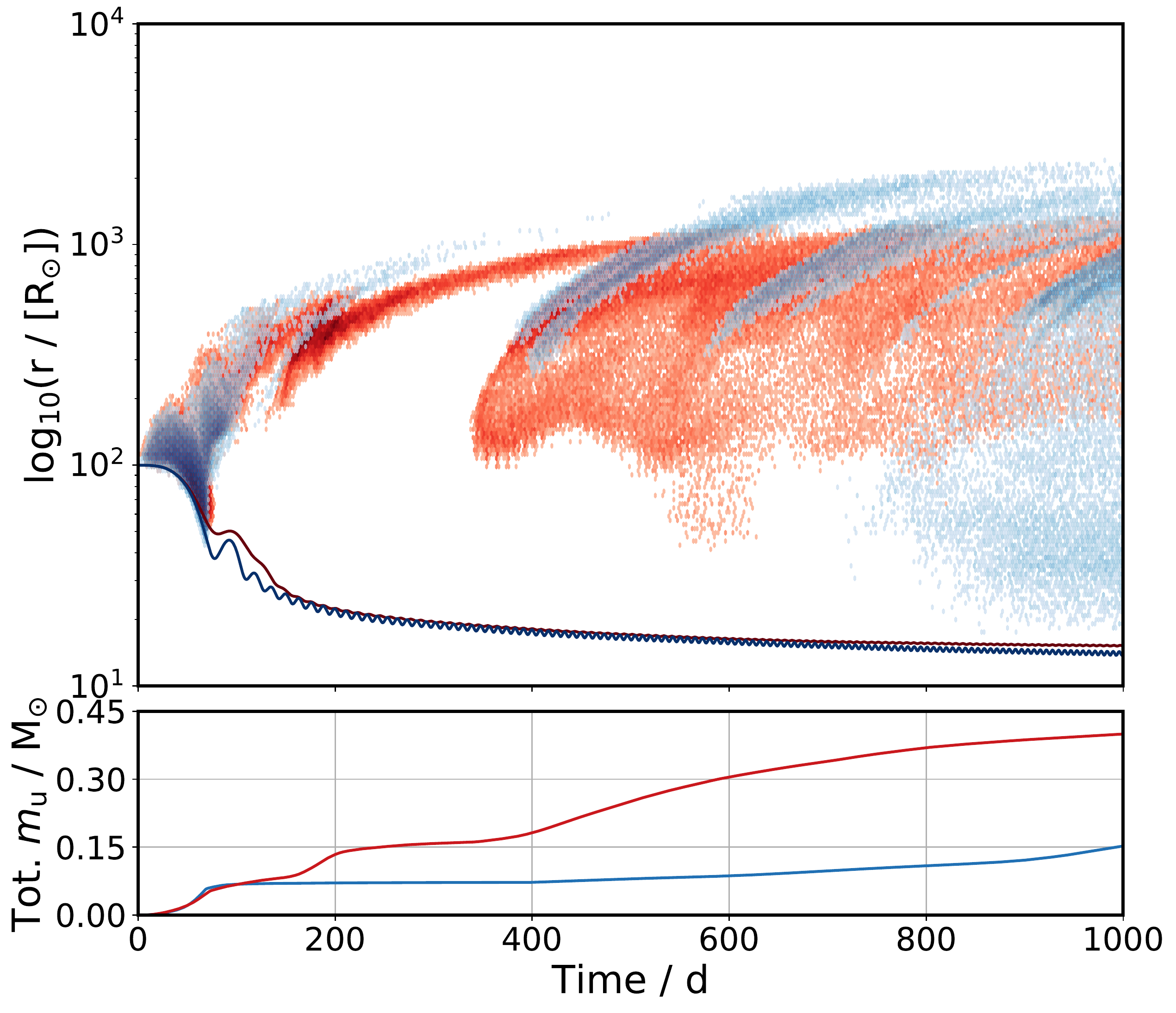}
    \hfill
    \includegraphics[width=0.49\linewidth]{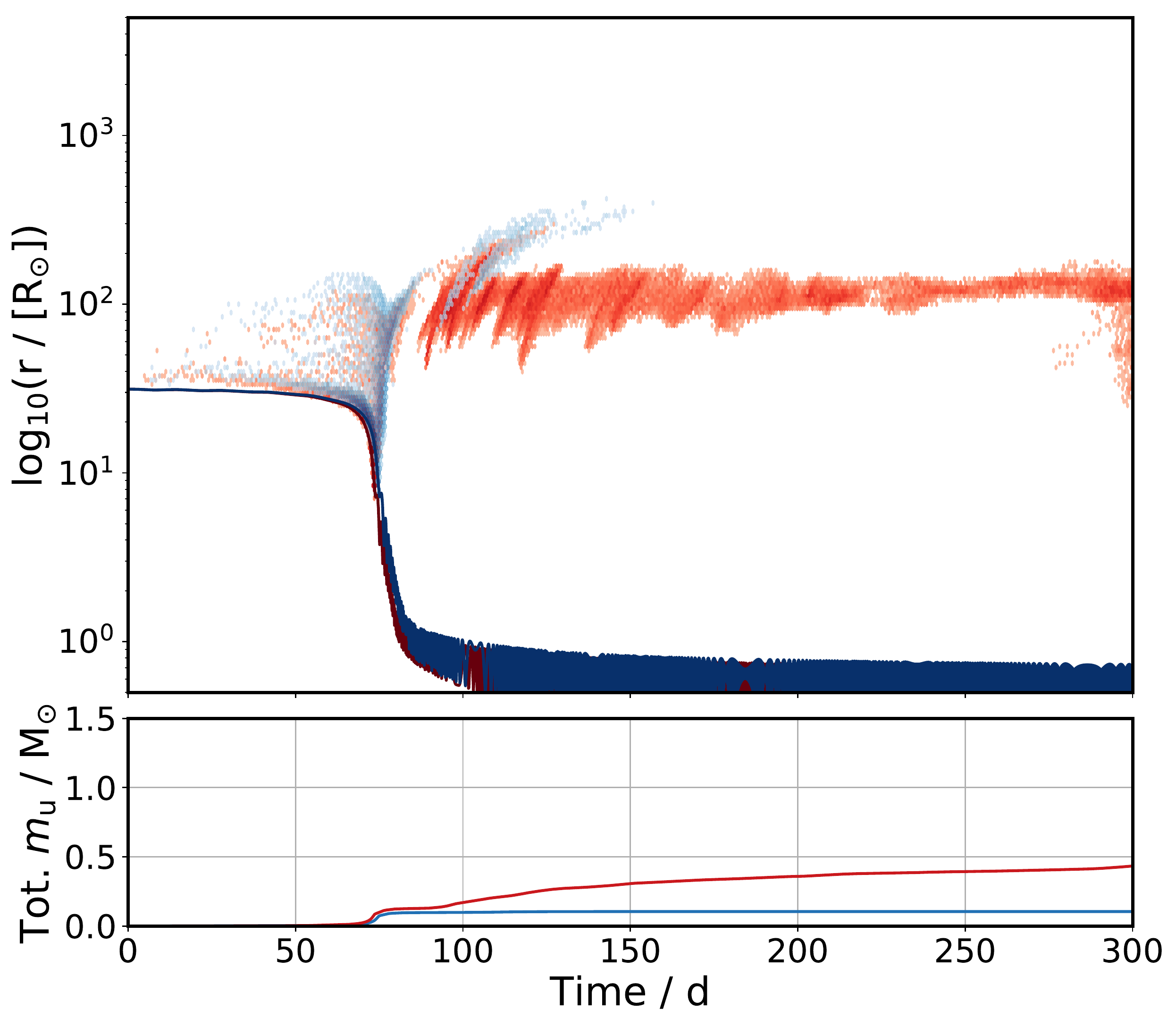}
    \caption{Radial location and time at which particles become unbound for the P12 (left) and N16 (right) simulations, showing the extra unbinding that takes place at late times for the two simulations using the tabulated equation of state (red) compared to the idea gas equation of state (blue). The particles are binned to show the relative strength of the unbinding events. In the main panels, the two lines represent the separation of the companion from the primary. The bottom panels give the cumulative sum of the unbound mass in the plot above.}
    \label{fig:unbound_particles}
    \includegraphics[width=0.49\linewidth]{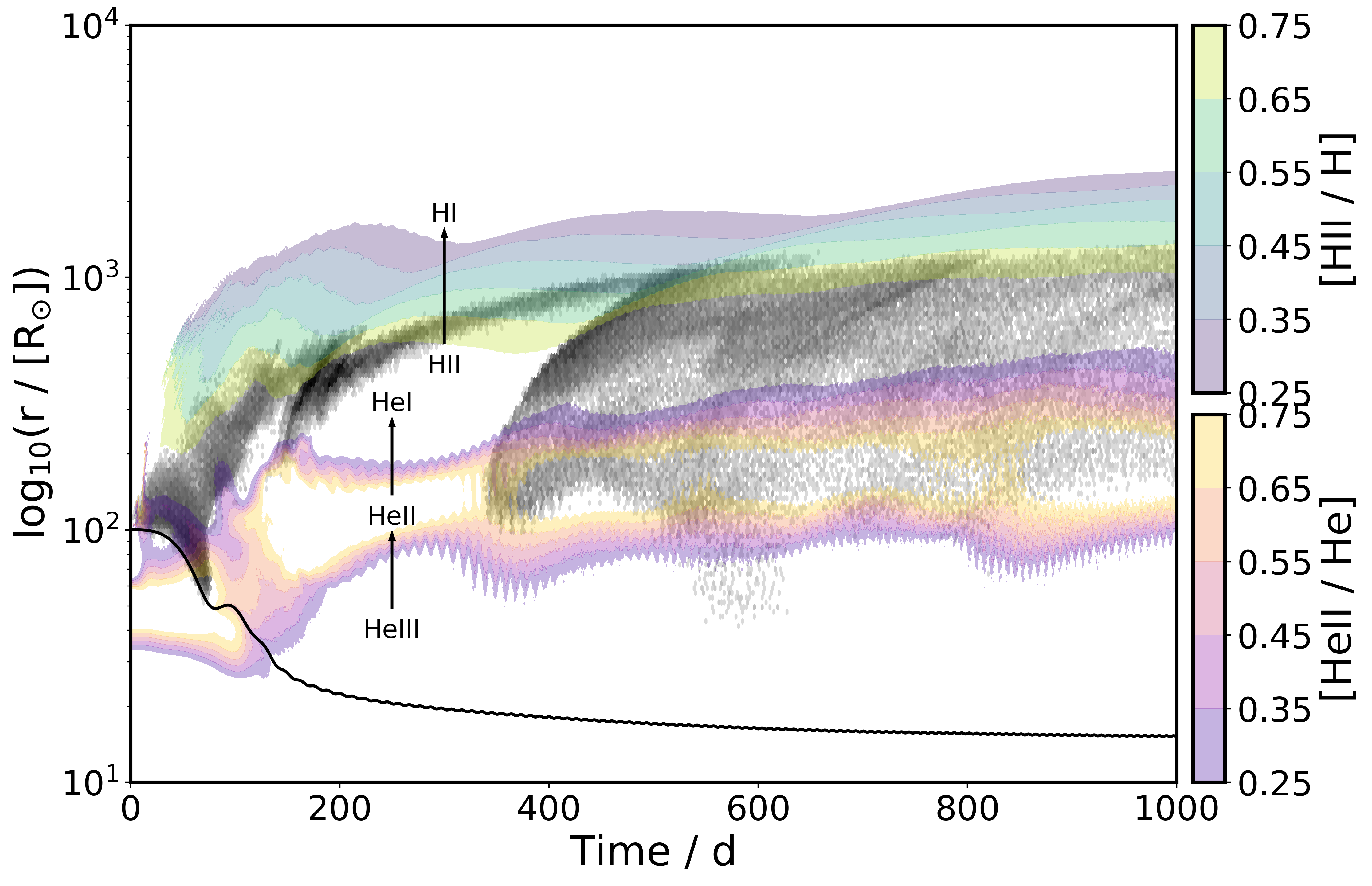}
    \hfill
    \includegraphics[width=0.49\linewidth]{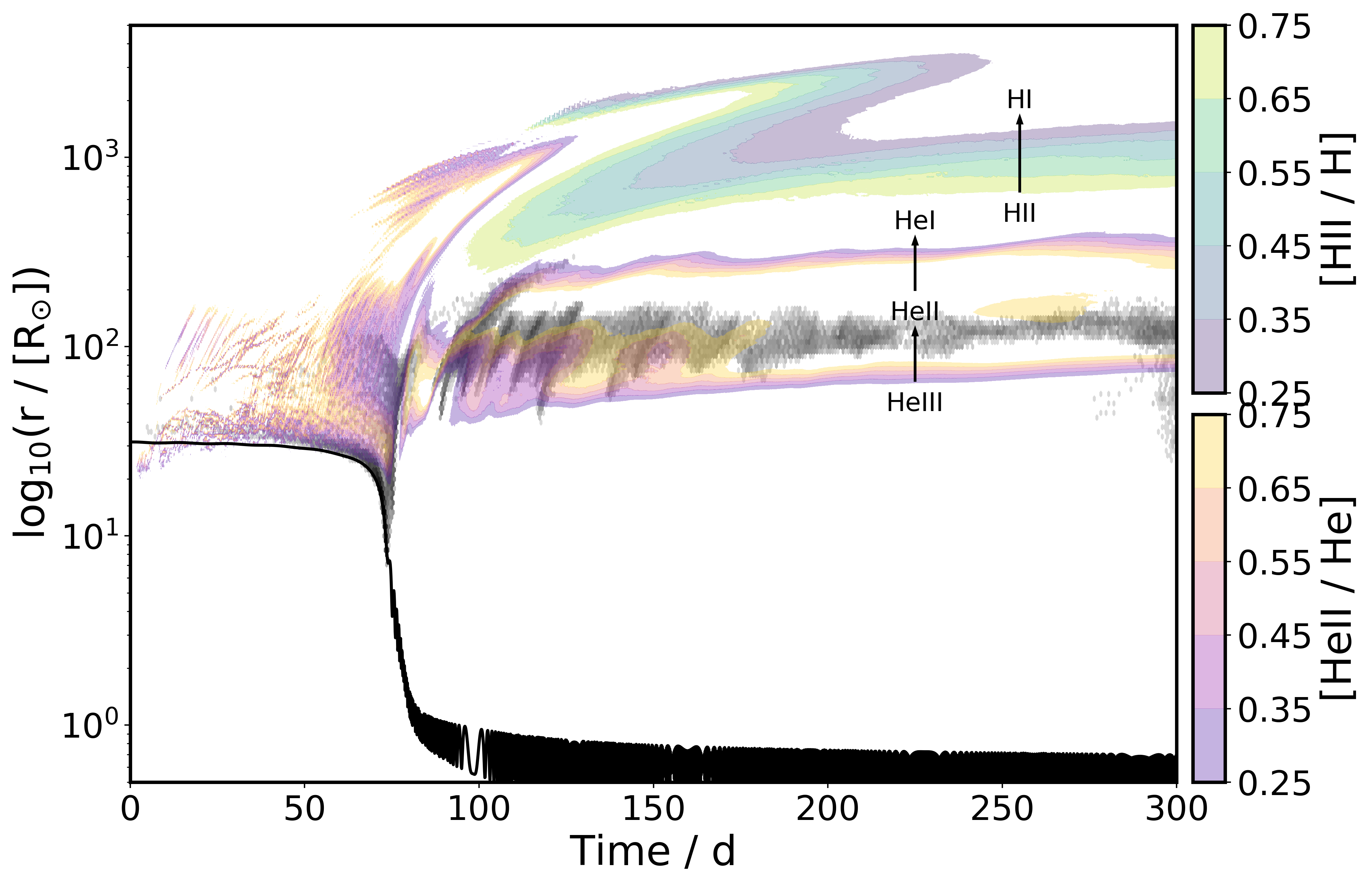}
    \caption{The correlation between unbinding of gas and recombination zones in the P12M (left) and N16M (right) simulations. In black, we overplot the locations and times at which particles are permanently unbound, that is, the red dots in Fig.~\ref{fig:unbound_particles}. The coloured zones represent the partially ionized regions for H and He, as seen in Fig.~\ref{fig:ions}.}
    \label{fig:unbinding}
\end{figure*}
Fig.~\ref{fig:ions} shows the radial distributions of HII and HeII in time in both the P12M and N16M simulations. Below the band of HeII in the bottom panels of Fig.~\ref{fig:ions}, the helium is entirely in the form of HeIII, and above the band it has fully recombined to neutral helium. To create these plots, we have calculated the ionisation fractions for each SPH particle in the simulation, and then averaged the resulting values in radial bins emanating from the core of the primary. We can clearly see that, in both systems, hydrogen begins to recombine in the envelope shortly after the beginning of the dynamic inspiral. However, as could be expected, the helium recombination zones do differ between the simulations. In the P12M simulation, the initial star has some helium recombination occurring near the outer layers, while the helium in the N16M star is initially completely in the form of HeIII. The N16M simulation has a short period of mass transfer before the fast inspiral, during which some gas particles are ejected from the system. In this ejected gas HeIII quickly recombines to HeII. In both simulations, recombination fronts of both hydrogen and helium form. After forming in the ejecta from the dynamic inspiral, these fronts remain at fairly constant radii from the central binary. Counter to some expectations, the zones do not appear to move inwards. Rather, gas flows outward from the central binary through these zones, recombining and releasing energy as it moves through the front. This release of recombination energy will be addressed in Section~\ref{ssec:recombination_release}.

\subsection{Where is the gas being unbound?}

Next we compared the locations where particles are being unbound with the recombination fronts seen in Section~\ref{ssec:where_recombination}. We call an SPH particle unbound by using criterion \textit{(ii)}, which is equivalent to criterion \textit{(iv)} in the ideal gas equation of state case (see Section~\ref{ssec:bound}). Criterion \textit{(ii)} includes only the mechanical energy and the thermal portion of the internal energy. We do not include  radiation energy, as our ideal equation of state does not include the effects of radiation pressure.

Fig.~\ref{fig:unbound_particles} shows the radial locations at which particles are being unbound in the simulations. These particles were then tracked backwards in time to determine the time and location at which they were last bound to the system. The bottom panels of these plots show the cumulative amount of permanently unbound material in the simulations. The common feature between the two panels of Fig.~\ref{fig:unbound_particles} is that all the simulations display a strong unbinding event at the beginning of the fast inspiral, which is driven primarily by the transfer of orbital energy to the gas. This particular feature shows up in all simulations of the common envelope, regardless of which equation of state is being used. It is visible in Fig.~\ref{fig:sep_bound} as a drop in the bound mass during the fast inspiral. However, in both the P12I and N16I simulations, the unbinding after the inspiral is quite weak, being limited primarily to the action of shocks propagating through the material. In comparison, both the P12M and N16M simulation display copious unbinding after the inspiral.

To determine the actual cause of the extra unbinding observed in the simulations that use our tabulated equation of state, we compare the location where particles become unbound to the recombination zones of hydrogen and helium. In Fig.~\ref{fig:unbinding}, we show the significant zones of partial ionization for both hydrogen and helium from Fig.~\ref{fig:ions}. We overplot the red points from Fig.~\ref{fig:unbound_particles} in greyscale. This reveals that, aside from the initial unbinding at the hand of orbital energy, the gas unbinding coincides with areas where hydrogen and helium are recombining.

In the P12M simulation (left-hand panel of Fig.~\ref{fig:unbinding}), the unbinding of particles is spatially coincident with the base of the hydrogen recombination zone and both helium recombination zones. However, for the N16M simulation (right-hand panel of Fig.~\ref{fig:unbinding}), unbinding occurs primarily in, or just above, the HeIII recombination zone. 

\citet{soker2018radiating} have argued that much of the hydrogen recombination energy is lost from the system through radiation and convection. Even if this were true, our simulations indicate that, for these setups, unbinding is promoted more by the recombination of helium than of hydrogen. Since the helium recombination zone is at a greater depth in the star, the energy released there is less likely to escape. Much of the gas experiencing hydrogen recombination has already been unbound at an earlier stage in these simulations, when it was deeper in the envelope of the primary. \textit{We suggest that for these simulations, it is the helium recombination energy that more strongly affects the unbinding of gas from the binary system.}

\subsection{Where is recombination energy being released?}
\label{ssec:recombination_release}

\begin{figure*}
    \centering
    \includegraphics[width=0.49\linewidth]{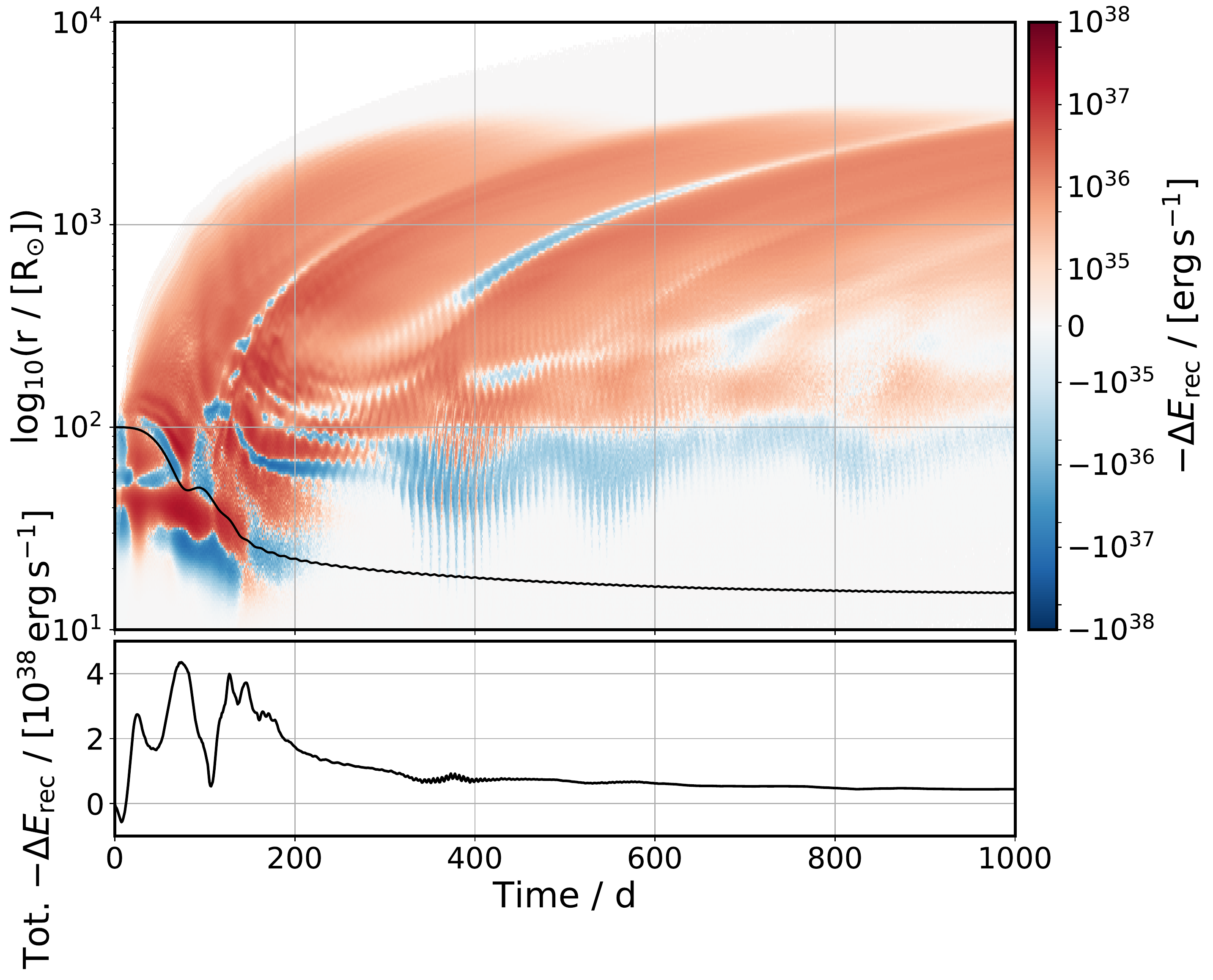}
    \hfill
    \includegraphics[width=0.49\linewidth]{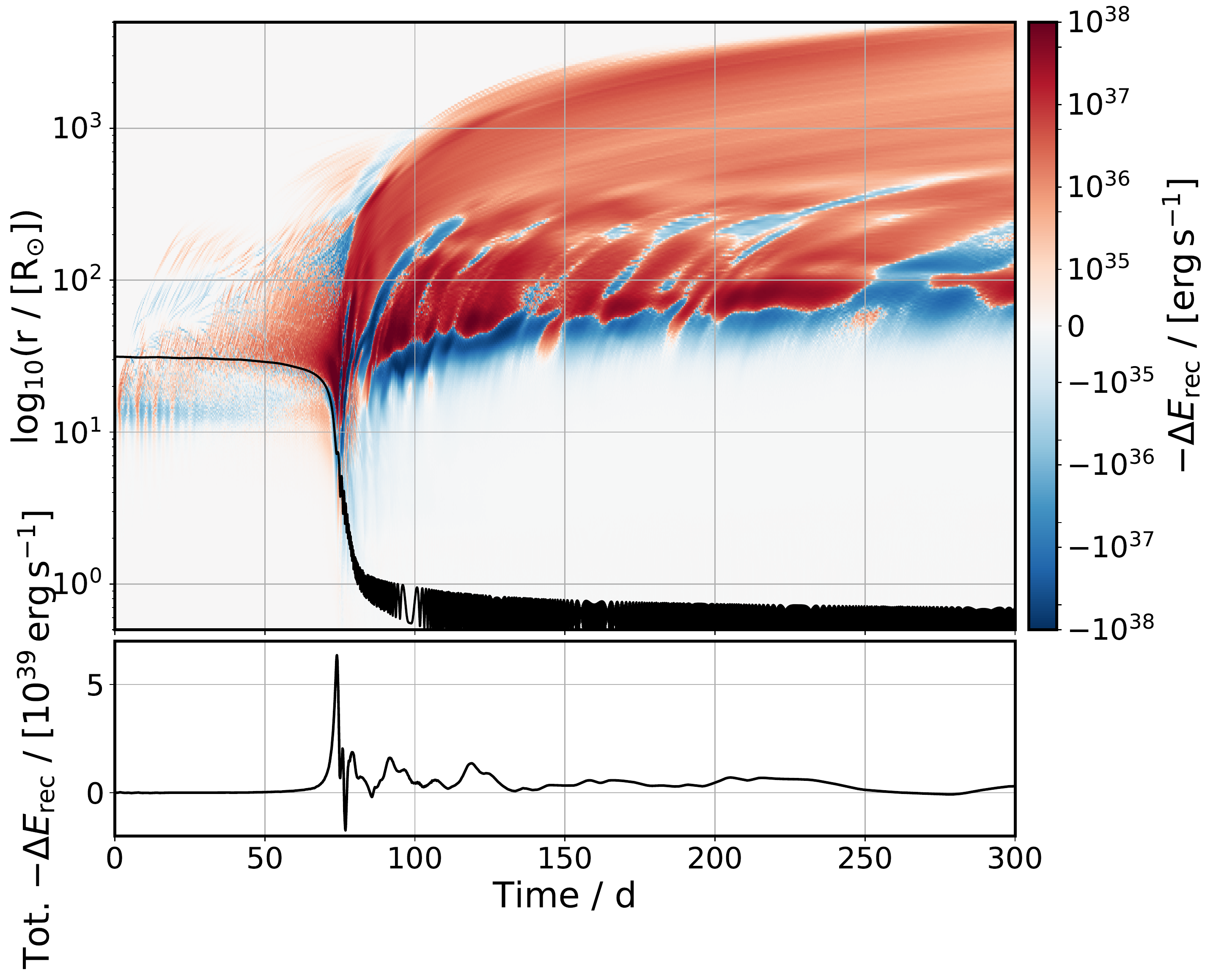}
    \caption{Recombination energy release rate in the P12M (left) and N16M (right) simulations, determined by plotting the change in recombination energy for each gas particle from one code output to the next. Positive numbers on the colour bar show that there has been a net decrease in the recombination energy of the gas, hence the energy has been released. The bottom panel is the total rate of recombination energy release at each time in the simulation.}
    \label{fig:erec}
\end{figure*}

In addition to knowing where the recombination fronts are, we can also calculate, using Eq.~\ref{eq:ereci}, where recombination energy is being released. Once again we radially bin the particles, and determine the change in recombination energy from the last code output. As our simulations have different lengths of time between code dumps, we divide the recombination energy delivered by the time between dumps, resulting in a rate of recombination energy release. These profiles are shown in Fig.~\ref{fig:erec}, where red areas show positive values, i.e., a net release of recombination energy, and blue areas represent negative values, i.e., particles being ionised. The lower panels of Fig.~\ref{fig:erec} show the net rate of recombination energy release at each moment of the simulation. The release of recombination energy per second is approximately an order of magnitude larger in the N16M simulation than the P12M simulation. This is likely due to the fact that there is more gas mass in the N16 simulations, and the ejection of gas during the inspiral is more rapid, causing gas to recombine more rapidly.

In Fig.~\ref{fig:erec} we see that, immediately after the inspiral, a clear boundary forms between areas where recombination energy is being released (red) and areas where it is being captured (blue) located at approximately the original orbital radius. This capture of recombination energy is due to the ionisation of HeII. It shows that gas is ionised initially close to the companion, from which shocks emanate moving outwards. In both simulations, this feature begins to dissipate quickly, reaching a point when the gas is no longer undergoing much ionisation, and recombination is only happening at larger radii.

There is a large release of recombination energy associated with the fast inspiral of the companion and the subsequent expansion of the envelope layers. In the P12M simulation, some of this energy release occurs inside the atmosphere of the giant star, just below the orbit of the companion. This may be due to the expansion of the innermost layers in response to the expansion of the layers above. This feature does not occur in the N16M simulation, which undergoes a less sudden inspiral due to its larger initial orbital separation. Further, the N16M star comprises only HeIII, hence the helium recombination is less sensitive to small changes in the stellar structure than the P12M star. However, after the initial inspiral, in both the P12M and N16M simulations, the recombination energy is released primarily above the orbit of the companion as the envelope expands and cools.

\subsection{Is the gas being unbound because of recombination energy?}

\begin{figure*}
    \centering
	\includegraphics[width=0.49\linewidth]{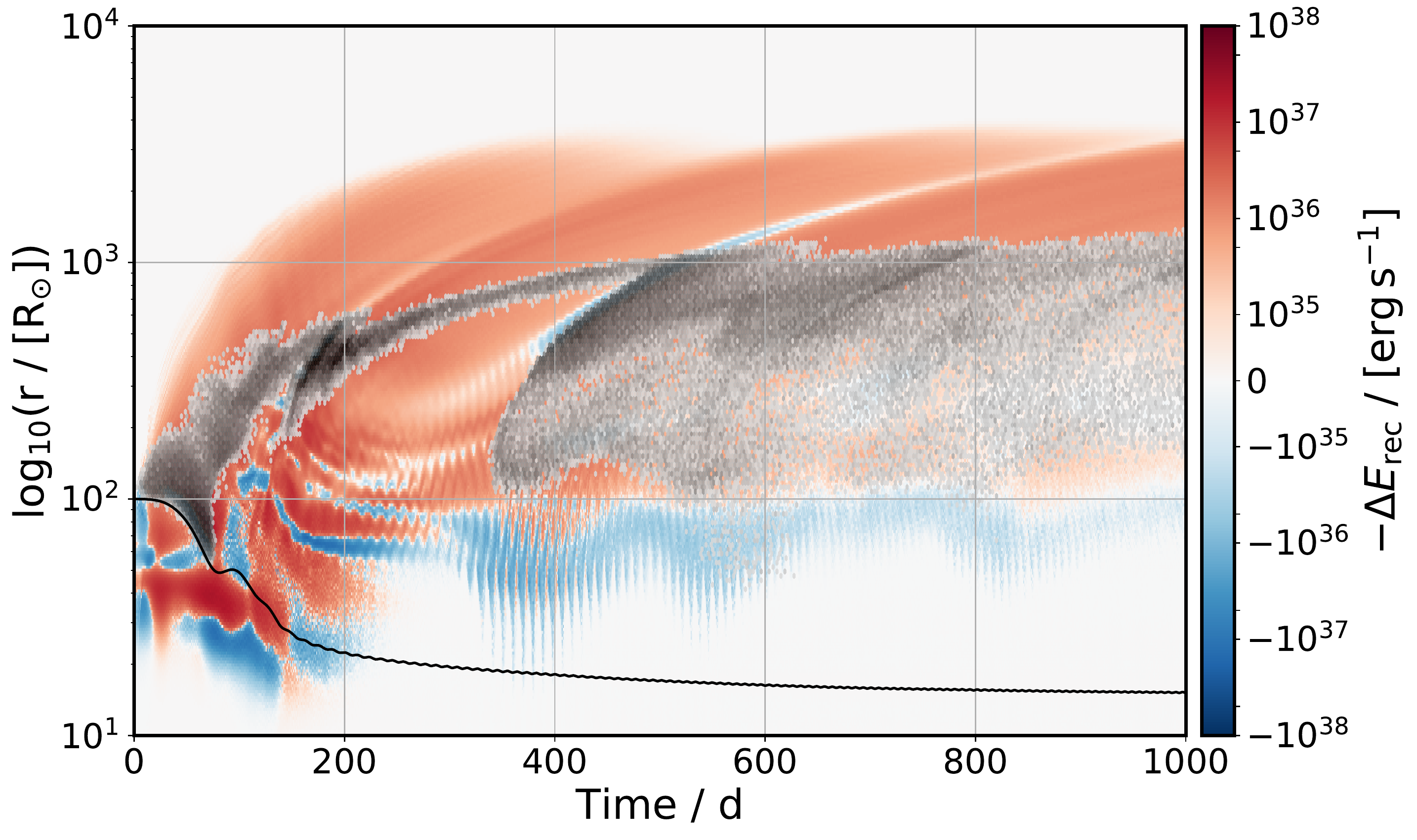}
	\hfill
	\includegraphics[width=0.49\linewidth]{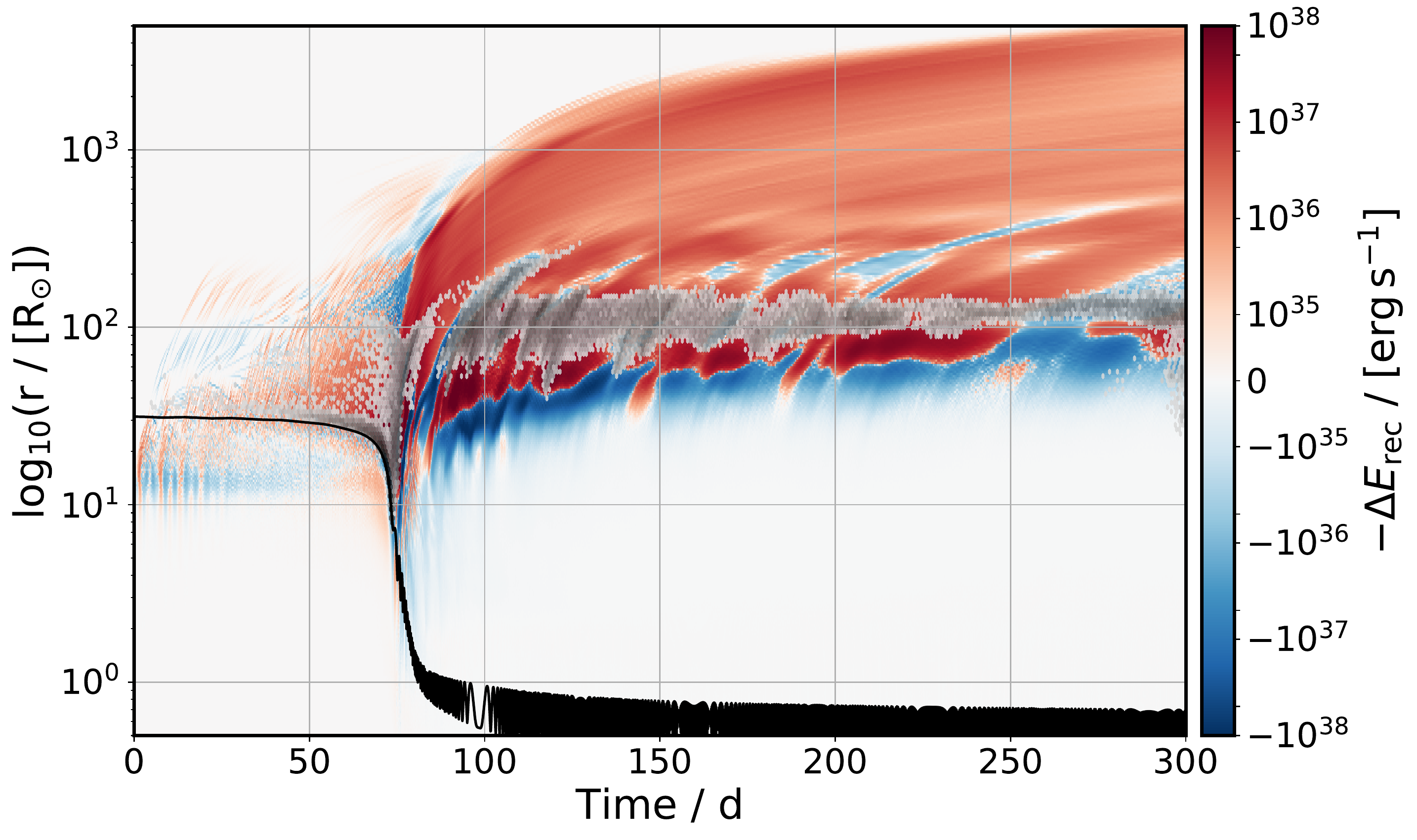}
    \caption{Released recombination energy in the N16M simulation, from Fig.~\ref{fig:erec}. The grey dots again represent the newly, permanently unbound particles from Fig.~\ref{fig:unbound_particles}.}
    \label{fig:unbound_erec}
\end{figure*}

Even if we can confirm that the tabulated equation of state simulations are unbinding more material, we cannot yet confirm that the recombination energy is the trigger for this extra ejection. To investigate this, in Fig.~\ref{fig:unbound_erec} we again plot the recombination energy being released in the simulation, and overplot in grayscale the particles that became unbound and remain unbound for the rest of the simulation, as was shown in Fig.~\ref{fig:unbound_particles}.

Looking at the left-hand panel of Fig.~\ref{fig:unbound_erec}, we can see that there are a couple of strong events where gas is being ionised (rendered in blue), suggesting that a shock is moving through the material. These are sites where we would expect particles to be unbound even without recombination, due to the increase in thermal energy of the gas as the shock moves through it. 

The release of recombination energy happens when the gas is expanding and cooling, as can occur in the wake of shocks. Without the extra recombination energy input, these gas particles would likely not be unbound in great numbers. This is observed in the left-hand panel of Fig.~\ref{fig:unbound_particles}, where the release of recombination energy for the P12M simulation enhances particle unbinding in these locations when compared to the P12I simulation. This suggests that the input of energy through shocks and recombination behind the shock may help to unbind the envelope.

On the other hand, the unbinding of particles in the N16M simulation aligns very closely with the strong release of helium recombination energy, as is presented in the right-hand panel of Fig.~\ref{fig:unbound_erec}. Contrasting with the unbinding that takes place in the N16I simulation (visible in Fig.~\ref{fig:unbound_particles}), where there is no extra input of energy, the particles are unbound almost exclusively during the inspiral. Clearly the helium recombination energy plays a strong role in unbinding this portion of the envelope.

\begin{table}
\begin{tabularx}{\linewidth}{Y >{\hsize=0.6\hsize}Y *{2}{>{\hsize=1.2\hsize}Y} >{\hsize=0.6\hsize}Y *{2}{>{\hsize=1.2\hsize}Y}}
\hline
Model & $t_\text{in}$ &  $\Delta E^\text{H}_\text{rec,in}$ &  $\Delta E^\text{He}_\text{rec,in}$ & $t_\text{f}$ &  $\Delta E^\text{H}_\text{rec,f}$ &  $\Delta E^\text{He}_\text{rec,f}$\\
 & (d) & ($10^{46}$\,erg) & ($10^{46}$\,erg) & (d) & ($10^{46}$\,erg) & ($10^{46}$\,erg) \\
\hline
P12M  &   359 &  0.23 &   0.34 &  1843 &  0.71 &  0.41 \\
N16M  &   158 &  0.09 &   0.45 &   922 &  0.56 &  1.61 \\
N16Mh &   138 &  0.15 &   1.20 &   553 &  1.31 &  1.70 \\
\hline
\end{tabularx}
\caption{Hydrogen and helium recombination energy ($\Delta E^\text{H}_\text{rec}$ and $\Delta E^\text{He}_\text{rec}$) released by 50\,d post inspiral ($t_\text{in}$) and the end of the simulation ($t_\text{f}$). }
\label{table:erec}
\end{table}

It appears that helium recombination energy may be sufficient to account for the extra unbinding in the tabulated equation of state simulations. In simulation P12M, helium recombination has released 3.4$\times 10^{45}$\,erg by 50\,d after the inspiral (Table~\ref{table:erec}). At the same time in the P12I simulation, 84~percent of the envelope remains bound. If we were to add the released helium recombination energy to maximise the number of unbound particles (i.e., to raise the energy of the least bound particles to zero by criterion \textit{(ii)}), we could reduce the bound envelope to 27~percent of the gas mass, which is significantly smaller than the 65~percent in the P12M simulation (Table~\ref{table:bound}). A similar calculation can be carried out for the N16 simulations. This would leave a bound envelope of 78~percent, which again is lower than the approximately 90~percent that is actually bound in the N16M simulation at the same point. In both cases, this maximum amount of unbinding is effectively impossible, as it implies that only the least bound particles are supplied with precisely enough energy to unbind them. However, it does show that the helium recombination can supply enough energy to account for the extra unbinding that takes place in the simulations that use our tabulated equation of state.

We note that, by the same reasoning, hydrogen recombination also delivers enough energy to fully account for the envelope unbinding. However, because of the stratification of the recombination zones, gas flowing through the hydrogen recombination zone would be already unbound by having passed through the helium recombination zone, and thus would likely have received the necessary energy to unbind it.

\section{The fraction of recombination energy that can do work}
\label{ssec:can_it_be_used}

In the previous sections, we have demonstrated that the inclusion of recombination energy in our simulations has caused an additional amount of envelope to become unbound. Our simulations are adiabatic and any released recombination photons are immediately thermalised into the gas, thus the energy is allowed to do work. Here we estimate the opacity of the gas in post-processing, and use it to calculate how much of the recombination energy would be radiated away, following the approach of \citet{grichener2018limited} and \citet{soker2018radiating}. However, we use quantities from our 3D simulations rather than relying on the 1D approach as they did.

\citet{grichener2018limited} and \citet{soker2018radiating}, expressed the photon diffusion time out of a recombination zone as \begin{equation}
    t_\text{diff} = \frac{3\tau\Delta R}{c},
    \label{eq:tdiff}
\end{equation}
where $\tau$ is the optical depth outwards from the recombination zone, $\Delta R$ is the depth of the recombination zone below the photosphere, and $c$ is the speed of light. \citet{soker2018radiating} also employed the convective timescale, which is defined instead as the timescale over which recombination photons may be removed from the common envelope by convection alone. Our simulations do not correctly capture the effects of convection, so we take instead the minimum possible convection timescale, given by:
\begin{equation}
    t_\text{min;conv} = \int^{R_\text{phot}}_{R} \frac{dr}{c_s(r)},
    \label{eq:tconv}
\end{equation}
where $R_\text{phot}$ is the radial location of the photosphere, $R$ is the radial location of the partial ionisation zone and $c_s(r)$ is the local sound speed. \citet{soker2018radiating} then defined the energy transport timescale as the minimum of the photon diffusion and convective timescales:
\begin{equation}
    t_\text{trans} = \min(t_\text{diff},t_\text{min;conv}).
\end{equation}
The third relevant timescale is the envelope expansion timescale, which is approximately on the order of the orbital timescale at the surface of the giant,
\begin{equation}
    t_\text{exp} \approx \frac{2 \pi R_1}{v_\text{kep}},
\end{equation}
where $R_1$ is the radius of the primary star and $v_\text{kep}$ is the Keplerian velocity at this radius. This timescale is important because, if the envelope expands on a similar timescale to $t_\text{trans}$, then more of the recombination energy will be utilised in the expansion of the envelope.

\citet{sabach2017energy}, \citet{grichener2018limited} and \citet{soker2018radiating} used these timescales to estimate the fraction of the recombination energy, $f_\gamma$, that can be used to accelerate the gas:
\begin{equation}
    f_\gamma < \left(\frac{t_\text{trans}}{t_\text{trans} + t_\text{exp}}\right).
    \label{eq:fgamma}
\end{equation}
The diffusion timescale increases as the star expands, hence $f_\gamma$ typically decreases over the course of the interaction. \citet{grichener2018limited} and \citet{soker2018radiating} continued this line of reasoning, using a 1D simulation calculated with the stellar structure code \textsc{mesa}. They evolved a 2\,\msun\ star until it was on the asymptotic giant branch (AGB), when it had a mass of 1.75\,\msun\ and a radius of 250\,\rsun. They then emulated a common envelope inspiral by injecting energy into the envelope, causing it to inflate from 250\,\rsun\ to about 520\,\rsun. This expansion happens over the course of 1.7 years, which they adopted as the envelope expansion timescale, $t_\text{exp}$. They discussed that from the zones where hydrogen ionisation is at about 30 per cent, the photon diffusion time $t_\text{diff} < 0.5$\,yr. With those values, \citet{soker2018radiating} estimated that, in their case, $f_\gamma < 0.2$.

In what follows, we use our simulations to make our own assessment of whether recombination energy unbinds the envelope, and whether it is realistic to use it to do work, as is done in codes with no radiation transport.

\subsection{The determination of the opacity}
\label{ssec:opacity}
\begin{figure}
    \centering
	\includegraphics[width=\columnwidth]{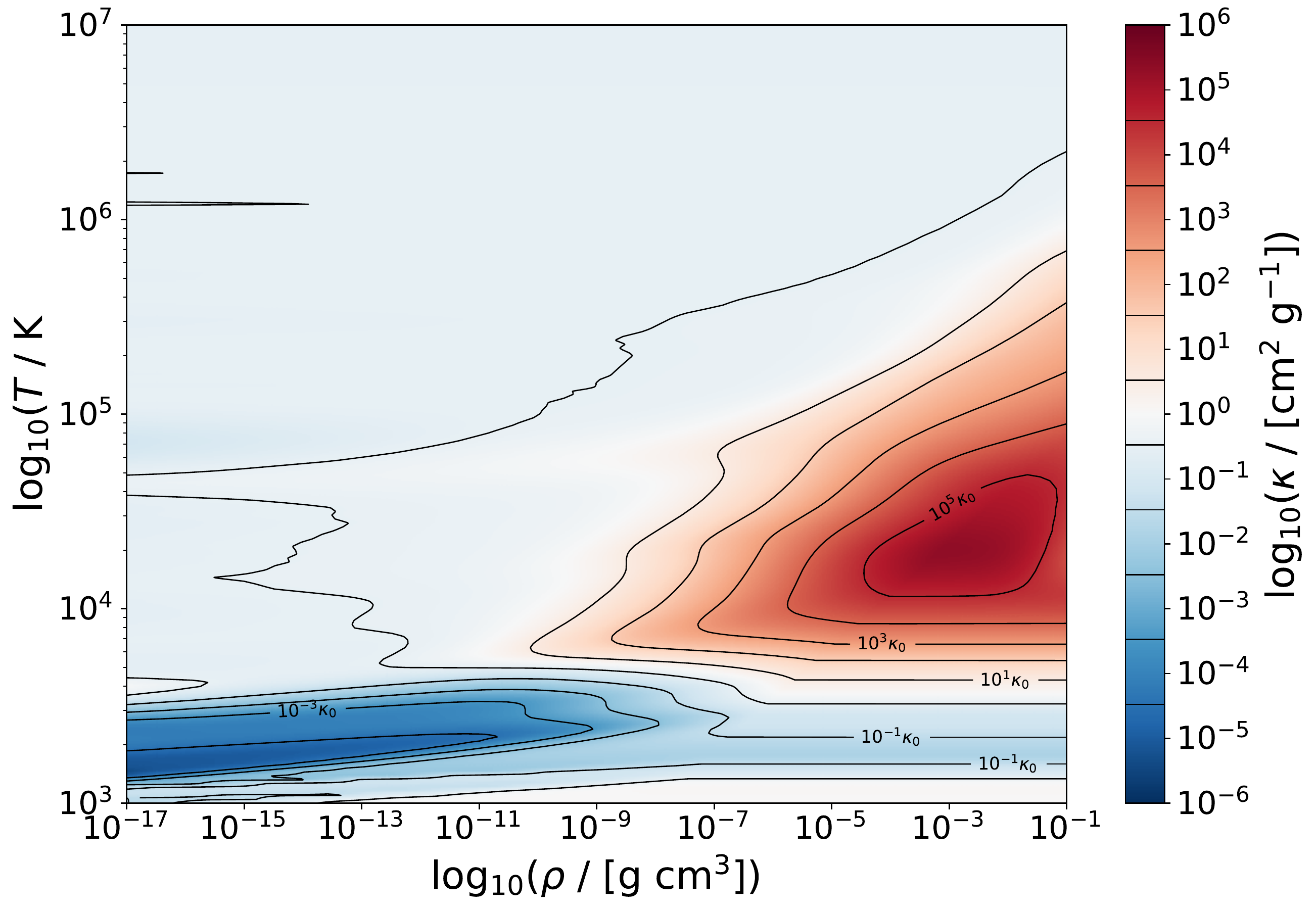}
    \caption{The opacity $\kappa$ as a function of temperature and density from the \textsc{mesa} opacity module. We plot the contours relative to the electron scattering opacity $\kappa_0 = 0.2(1+X)$\,cm$^2$\,g$^{-1}$, which is dominant in low density, low temperature, fully ionised gases. These contours are plotted to reflect the plot of opacity given in figure~3 of \citet{paxton2011modules}.}
    \label{fig:mesa_kappa}
\end{figure}

Alongside the \textsc{eos} module, \textsc{mesa} also has an opacity module, \textsc{kap}, which returns values of $\kappa$, given an input density and temperature. Like the equation of state data, the opacity data are drawn from several different sources, combining electron conduction and radiative opacities. The electron conduction opacities are given primarily by \citet{cassisi2007updated}, while the radiative opacities are given by tables from \citet[][for $2.7 \geq \log T \geq 4.5$]{ferguson2005low} and \citet[][for $3.75 \geq \log T \geq 8.7$]{iglesias1993radiative, iglesias1996updated}. For temperatures $\log T > 8.7$, Compton scattering dominates the opacity and is calculated with the equations of \citet{buchler1976compton}, while the low temperature tables of \citet{ferguson2005low} include the effects of molecules and grains on the opacity. 

The \textsc{mesa} opacity tables have been formatted so as to be able to be queried by \textsc{phantom}, with inputs of $X$, $Z$, $\rho$ and $T$, the last of which is an output from the equation of state itself. In principle, this can be used to estimate the photon diffusion timescale using Eq.~\ref{eq:tdiff}, similarly to what was done by \citet{sabach2017energy}, which requires the optical depth between the recombination zone and the photosphere.

While it would be more appropriate to calculate, during the simulations, the photosphere of the gas, as well as the optical depth of the location of SPH particles where recombination energy is being released, neither of these tasks are trivial. For this reason, we use the opacity data purely in our analysis of the tabulated equation of state simulations, emphasising that it was not used during the simulations.

\subsection{The determination of timescales}
\begin{figure*}
    \subfloat[P12M profiles, taken at $t=358$\,d \label{fig:p12_profiles}]{\includegraphics[width=0.49\linewidth]{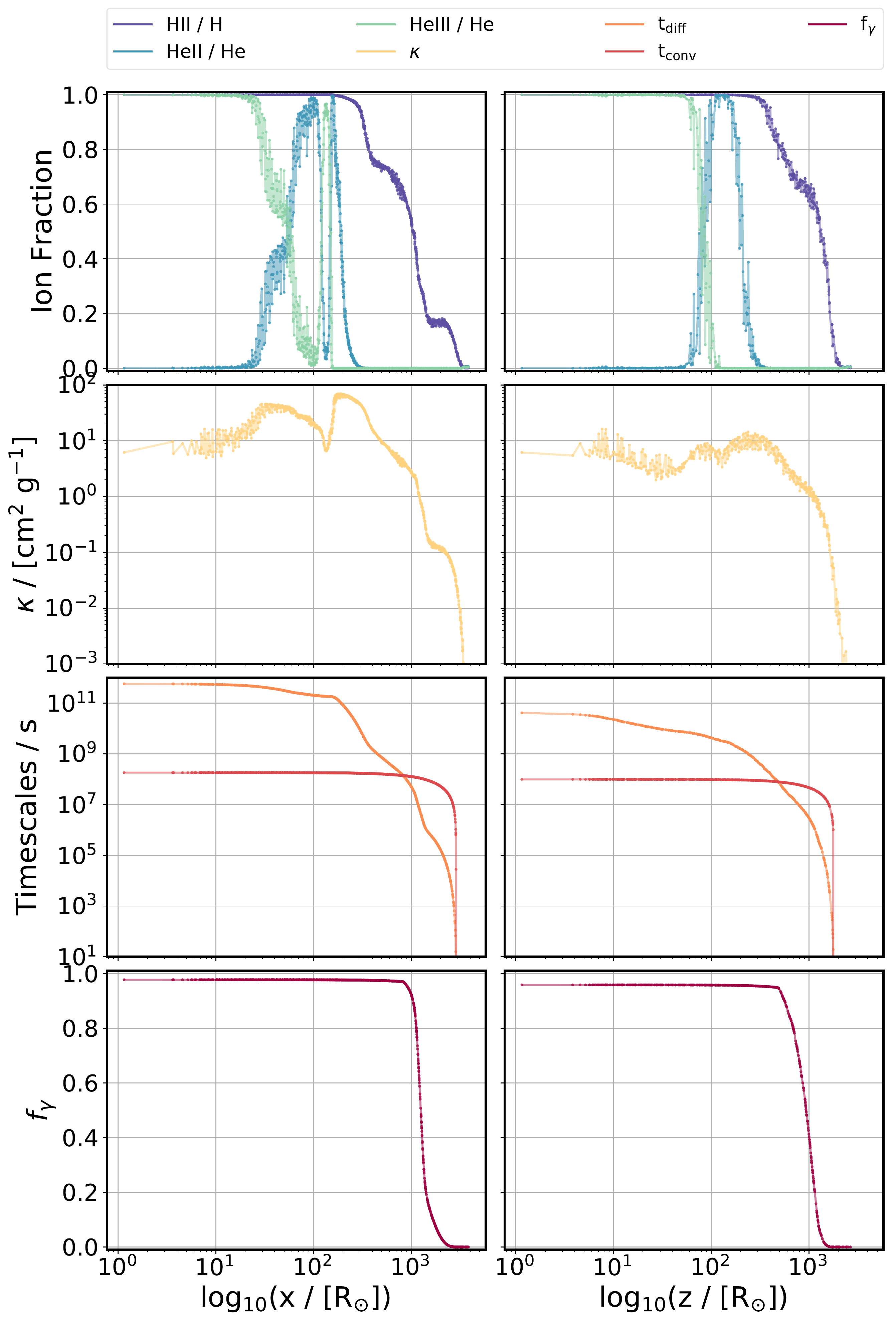}}
    \hfill
    \subfloat[N16M profiles, taken at $t=157$\,d \label{fig:n16_profiles}]{\includegraphics[width=0.49\linewidth]{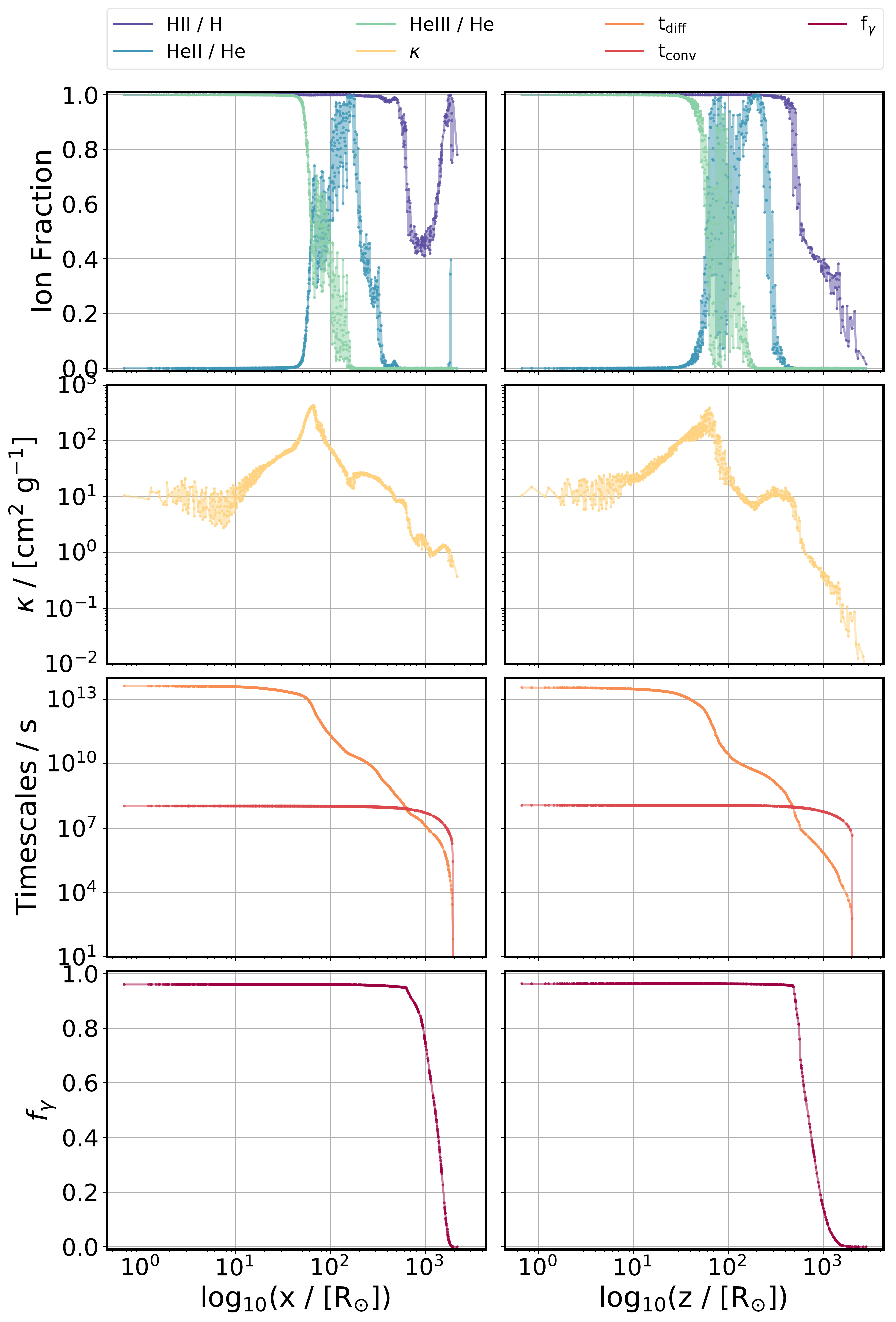}}
    \caption{Profiles of the P12M and N16M simulations, which is 50\,d after the end of their respective fast inspirals. Left columns: Profiles are taken in the positive x-direction, from the primary's core. Right columns: Profiles are taken in the positive z-direction, from the primary's core. First row: Ionisation fractions for hydrogen (HII/H) and helium (HeII/He and HeIII/He). Second row: $\kappa$ returned from the \textsc{mesa} opacity tables. Third row: Comparison between the photon diffusion timescale, $t_\text{diff}$, and the minimum convective timescale, $t_\text{min;conv}$. Fourth row: Approximately derived profile for $f_\gamma$. Each point represents one particle, which are joined to help see the trends in the data.}
\end{figure*}
The optical depth within the envelope is determined by performing the integration \begin{equation}
    \tau = \int_{r_1}^{r_2} \kappa(r)\rho(r) dr,
\end{equation}
where $\tau$ is the optical depth, $\kappa$ is the opacity, $\rho$ is the density and $r_1$ and $r_2$ are the locations between which the optical depth is to be calculated. We find the photosphere by determining the location where the optical depth $\tau \approx \frac{2}{3}$, when integrating in from the outside. The timescale of envelope expansion is determined  by measuring the size of the photosphere averaged over the three axes. We find that the envelope approximately doubles in size within about 50\,d after the beginning of the fast inspiral, in both the P12M and N16M simulations. Since the radial locations of recombination zones were already determined in Section~\ref{ssec:where_recombination}, we  have all the necessary information to estimate $f_\gamma$, the fraction of recombination energy that can be used to do work, as defined in Eq.~\ref{eq:fgamma} \citep{soker2018radiating}.

Figs~\ref{fig:p12_profiles} and \ref{fig:n16_profiles} show profiles for the P12M and N16M simulations, respectively. These profiles radiate from the centre of mass of the simulation, at 50\,d after the end of the dynamic inspiral, in the $x$ and $z$-directions (left and right columns, respectively). The top row shows ionisation fractions for hydrogen and helium, the second row gives values of $\kappa$ from the \textsc{mesa} opacity tables, the third row is a comparison between the photon diffusion timescale $t_\text{diff}$ and the minimum convective timescale, $t_\text{min;conv}$ and the bottom row shows $f_\gamma$. As mentioned previously, we used an estimated envelope expansion timescale of about 50\,d for both simulations.

We can see from the bottom row that, for much of the inner portion of the simulation, the majority of the released radiation can be used in helping to eject the envelope. There are a couple of effects to consider here. Firstly, our simulations do not display convection, so we must use the minimum convective timescale as our estimate for how much radiation can be removed. That is, we must assume that convection is able to remove the recombination energy at the maximum rate, to obtain a lower limit on how much recombination energy may be used for envelope expansion. But, as was explained by \citet{ivanova2018use}, convection cells in giant stars typically operate at highly subsonic velocities, while the minimum convective timescale implies that convection cells are moving at approximately the local sound speed. Hence, we can say that either the convective timescale is longer in real systems, or as stated by \citet{ivanova2018use}, some energy would be used to accelerate the convection cells to nearly sonic velocities. This suggests that $f_\gamma$ may actually be higher than is shown in Figs~\ref{fig:p12_profiles} and \ref{fig:n16_profiles}, at least in the inner regions before the diffusion timescale takes over.

In both simulations, the recombination zone for hydrogen approximately coincides with the zone where $f_\gamma$ begins to drop sharply. This is due to the fact that the opacity plummets when the gas particles are fully recombined. Given that the hydrogen recombination zone straddles the region where $f_\gamma$ drops from about 0.95 to 0, we crudely estimate that only about half of the hydrogen recombination energy is usable, that is, $f_\gamma \approx 0.5$ for hydrogen. This estimation suggests that only a half of the hydrogen recombination energy may be used to expand the envelope. 

On the other hand, there is a strong peak in the opacity which aligns with the partial ionisation zones of HeII and HeIII. The picture is much clearer for helium. In the zones where helium recombination energy is being released, $f_\gamma \sim 0.95$. This value is purely driven by the \textit{minimum} convective timescale. Therefore, we estimate that more than 0.95 of released helium recombination energy can be utilised to eject the envelope, particularly if the convective timescale is longer than we have calculated here.

\section{Possible dusty shell formation}
\label{ssec:dusty_shell}
\begin{figure*}
	\includegraphics[width=0.75\linewidth]{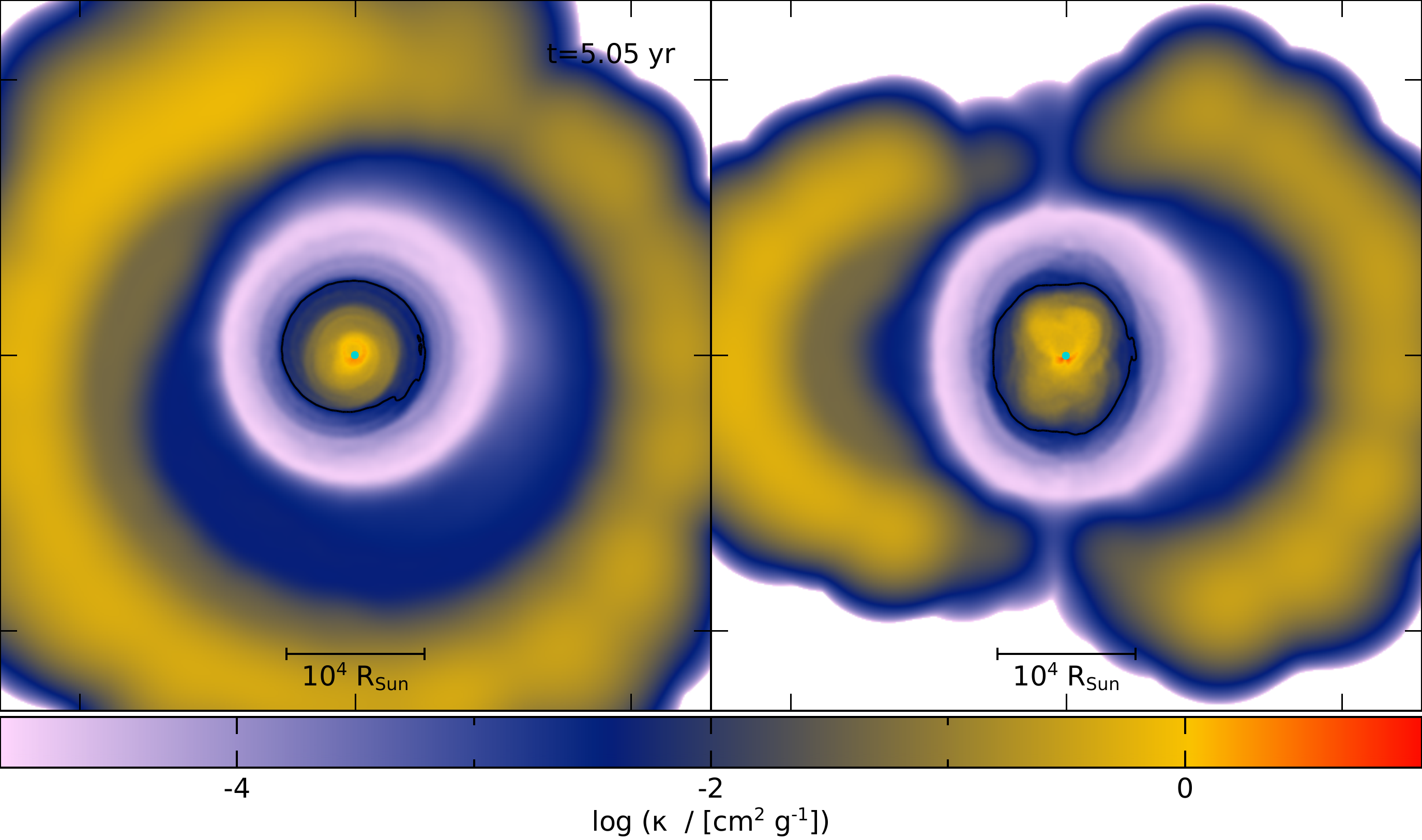}
	\includegraphics[width=0.75\linewidth]{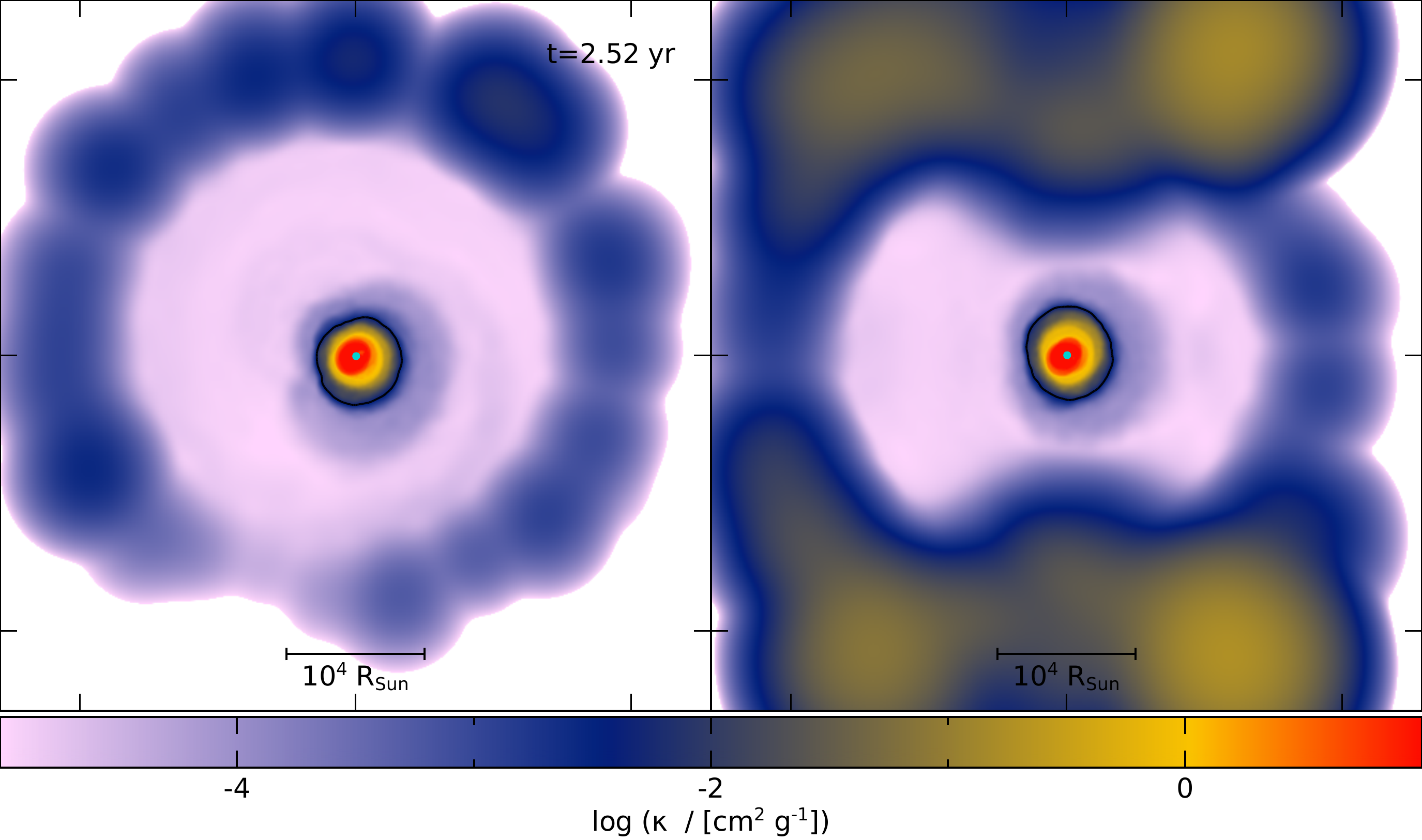}
    \caption{Rendering of $\kappa$ in the xy-plane (left) and in the xz-plane (right) of the P12M simulation (top) and the N16M simulation (bottom). The snapshots are taken at the end of the simulations at $t=5.05$\,yr and $t=2.52$\,yr for the P12M and N16M simulations, respectively. The black contour shows the boundary outside of which all hydrogen has recombined.}
    \label{fig:kappa}
\end{figure*}
In Fig.~\ref{fig:kappa} we plot a rendering of the opacity, $\kappa$, at the end of the P12M and N16M simulations ($t=5.05$\,yr and $t=2.52$\,yr, respectively). In the P12M simulation, we find that the central, high opacity envelope is surrounded by a very low opacity shell, just outside of the black contour, where all hydrogen is fully recombined. However, an interesting feature is also visible. Just beyond this low opacity shell, a higher opacity shell begins to develop from about $t=4$\,yr into the P12M simulation and $t=2$\,yr into the N16M simulation. These times correspond to about 3\,yr and 1.6\,yr after the dynamic inspirals of the P12M and N16M simulations, respectively. We expect that these shells would continue to develop if our simulations were continued for longer.

At the end of the P12M simulation, we estimate that there is 0.19\,\msun\ outside the neutral hydrogen contour (visible in the top panels of Fig.~\ref{fig:kappa}). Of this 0.19\,\msun, approximately 0.06\,\msun\ resides in the high opacity shell. In the N16M simulation (bottom panels of Fig.~\ref{fig:kappa}), there are 0.14\,\msun\ outside the neutral hydrogen contour, 0.08\,\msun\ of which is beginning to form a similar high opacity shell. The high opacity shells extend between $\sim$20\,000 and $\sim$26\,000\,\rsun\ ($\sim$93--120\,AU) from the binary by the end of the P12M simulation, and between $\sim$18\,000 and $\sim$21\,000\,\rsun\ ($\sim$ 83--97\,AU) from the binary by the end of the N16M simulation. 

The physical reason for the increase in opacity is that dust begins to form in the low temperature regions near the outer boundary of the expanding gas. In particular, the opacity tables by \citet{ferguson2005low} supply, for a certain combination of temperature and density, an average opacity calculated from several types of molecules and grains.

In our simulations we do not actually include the effects of dust, therefore the dynamics and thermodynamics of the expanding envelope are not affected by the formation of dust. The high opacity regions are instead inferred from the gas temperature, density and composition in post-processing of the simulations (similarly to what was done by \citealt{iaconi2019properties}).

Two interesting possibilities are opened by this discovery. The first is that whatever energy leaks out of the photosphere (which in simulations like ours is located just outside the hydrogen recombination zone) will likely be intercepted by the opaque shell and possibly result in accelerating it. The ``dusty" region only contains a fraction of the envelope gas, but this fraction progressively increases as the simulation continues.

Second, the morphology of the high opacity shell is not spherical, with regions of much lower opacity forming radiation holes. Curiously the P12M simulation has lower opacity regions at the poles, while the N16M simulation has lower opacity regions at the equator. If radiation is intercepted by this high opacity material and does accelerate it, it will do so unevenly, likely partaking in the shaping of the envelope. This conclusion may be very relevant for planetary nebula formation and shaping and add to the complexity of shaping post-CE planetary nebulae already studied by \citet{garcia2018common} and \citet{frank2018planetary}. 

\section{Summary and conclusions}
\label{sec:mesa_conclusions}
We have presented a comparison of simulations carried out with ideal gas or tabulated equations of state to quantify the effects of recombination energy on the common envelope binary interaction. We have tested two different primary stars both of which were previously modelled. These simulations are still far from full radiation transport models, but they offer valuable insight regardless. We conclude the following:

\begin{enumerate}
    \item The final orbital separation is not influenced by the equation of state choice. We conclude that extra unbinding observed in simulations that release recombination energy does not impact the orbital inspiral. The orbital energy is typically injected earlier and at smaller radii than where gas recombines, leaving the inspiral largely unchanged. This conclusion is in line with what concluded by \citet{nandez2016common}.
    
    \item A far greater amount of envelope gas is unbound when our tabulated equation of state is used. This is more so for the 0.88\,\msun\ giant simulation, where effectively the entire envelope is unbound, than for the 1.8\,\msun\ one. The criterion adopted to determine whether gas is unbound is important. We advocate the use of a criterion that includes the thermal energy of the gas, but would suggest that a pure mechanical energy criterion be always used as a comparison. Like the simulations of \citet{nandez2015recombination} and of \citet{nandez2016common}, all recombination energy is converted to work in our adiabatic simulations.
    
    \item The fraction of hydrogen recombination energy available to unbind the envelope has been recently debated. We add incremental (though not definitive) evidence that suggests that the amount of released hydrogen recombination energy available should be of the order of 50~percent. 
    
    \item However, we also show that helium recombination energy is very effective in unbinding gas because virtually none of it can escape. For the case of our 1.8\,\msun\ simulation, it is the helium recombination energy that dominates the envelope unbinding, while for our 0.88\msun\ giant simulation helium contributes approximately half of the recombination energy. It is curious how relatively similar stars exhibit substantially different behaviours. This cautions us against generalising these conclusions to other intermediate mass stars. 
    
    \item Dust is observed to form in the remnants of common envelope interactions \citep{nicholls2013dusty}. We confirm that in the outer layers of expanding common envelopes the conditions are such that dust opacity dominates. In both our simulations we observe a high opacity structure, which we interpret as a dusty shell with prominent axi-symmetric morphologies. It is therefore likely that any energy that crosses the photosphere will be captured within this dusty shell, which contains a few tens of solar masses of gas, and possibly be able to accelerate it. 
\end{enumerate}

\section*{Acknowledgements}

TR acknowledges financial support from the Macquarie University Research Excellence scholarship. TR and OD acknowledge support from the Monash Centre for Astrophysics during several visits. OD and DP acknowledge funding via  Australian Research Council Future Fellowships FT120100452 and FT130100034, respectively. RI acknowledges the financial support provided by the Postodoctoral Research Fellowship of the Japan Society for the Promotion of Science (JSPS P18753). The team is grateful to Jose Nandez and Natasha Ivanova for providing the stellar structure used in their work.




\bibliographystyle{mnras}
\bibliography{mnras_bib} 


\appendix
\section{Resolution test}
\begin{figure*}
    \centering
	\includegraphics[width=\linewidth]{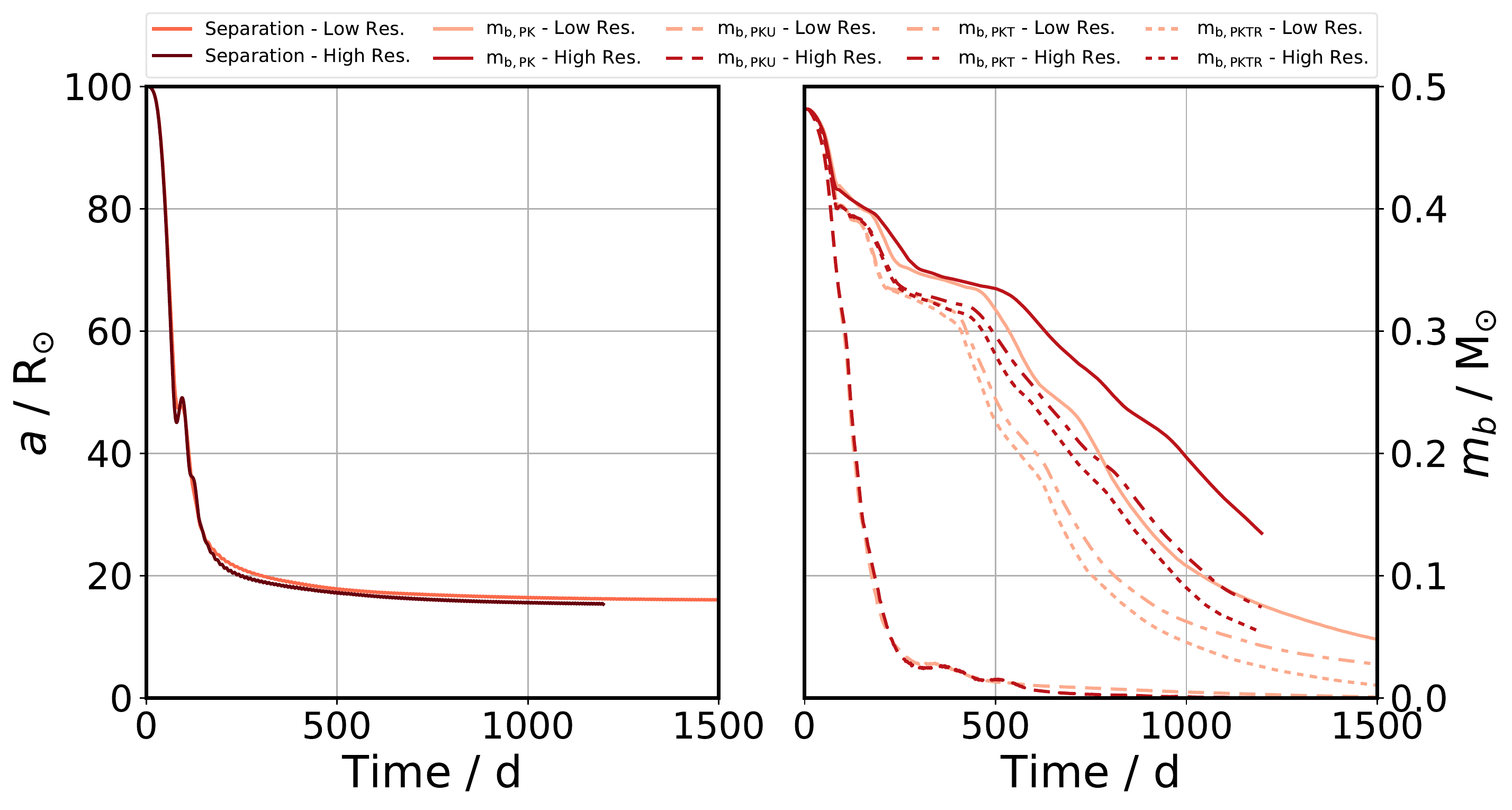}
    \caption{Comparison of orbital separation (left) and bound mass (right) for low resolution ($8\times 10^4$~particles; blue) and high resolution ($8\times 10^5$~particles; red) P12M simulations. The bound mass labels are the same as Fig.~\ref{fig:sep_bound}.}
    \label{fig:p12_sep_bound_old}
\end{figure*}

We carried out the P12M simulation at higher resolution using $8\times 10^5$ SPH particles to study the effect of resolution. This simulation was executed using a slightly older version of \textsc{phantom} and a slightly less refined stellar stabilisation method. We have not repeated it with the updated code because of a large wall clock time. We expect there to be negligible differences between outputs, but we have nonetheless compared this high resolution simulation to a low resolution simulation ($8\times 10^4$ SPH particles) carried out with the same code version.

Fig.~\ref{fig:p12_sep_bound_old} shows how the orbital separation and bound mass evolve with time for these two resolutions. The evolution of the bound mass, using our four criteria, remain within 10-20~percent within the first 500\,d, after which resolution effects become larger. This suggests that the extra unbinding observed in simulations including recombination energy is not a resolution effect, and that the recombination energy is indeed being used to eject the envelope.

A ten-fold increase in SPH particle number (corresponding to approximately a doubling of linear resolution) does not alter the behaviour of our results and the conclusions of our study.


\bsp	
\label{lastpage}
\end{document}